\documentclass[twocolumn,aps,prl,superscriptaddress,notitlepage,longbibliography]{revtex4}
\usepackage[colorlinks=true,urlcolor=blue,citecolor=blue,linkcolor=blue]{hyperref}
\usepackage{graphicx}  
\usepackage{dcolumn}   
\usepackage{bm}        
\usepackage{amssymb}
\usepackage{amsfonts}
\usepackage{doi}
\usepackage{verbatim}
\usepackage{epstopdf}

\usepackage{amsmath}
\usepackage{braket}
\usepackage{hyperref}
\DeclareMathOperator{\Tr}{Tr}

\begin{document}

\title{ Hinged Quantum Spin-Hall Effect in  Antiferromagnetic Topological Insulators}
\author{Yue-Ran Ding}
\affiliation{Institute for Advanced Study and School of Physical Science and Technology, Soochow University, Suzhou 215006, China.}
\author{Dong-Hui Xu}
\affiliation{Department of Physics, Hubei University, Wuhan 430062, China. }
\author{Chui-Zhen Chen}\thanks{czchen@suda.edu.cn}
\affiliation{Institute for Advanced Study and School of Physical Science and Technology, Soochow University, Suzhou 215006, China.}
\author{X. C. Xie}\thanks{xcxie@pku.edu.cn}
\affiliation{International Center for Quantum Materials, School of Physics, Peking University, Beijing 100871, China}
\affiliation{CAS Center for Excellence in Topological Quantum Computation,
University of Chinese Academy of Sciences, Beijing 100190, China}
\affiliation{Beijing Academy of Quantum Information Sciences, West Bld.3,
No.10 Xibeiwang East Rd., Haidian District, Beijing 100193, China}
\begin{abstract}
 In this work, we predict a hinged quantum spin-Hall (HQSH) effect featured by a pair of helical hinge modes in antiferromagnetic (AFM) topological insulator (TI) multilayers.
 This pair of helical hinge modes are localized on the hinges of the top and bottom surfaces of the AFM TI multilayers. Unlike the conventional QSH effect, the HQSH effect survives the breaking of time-reversal symmetry (TRS) and thus represents a different kind of topological phenomenon. The helical hinge modes are sustainable to inelastic scattering and TRS-breaking disorder, which can be observed in macroscopic samples. We show that this HQSH effect can be understood as a three-dimensional generalization of the Su-Schrieffer-Heeger model and its topology is characterized by the spin Chern number. At last, we propose that the HQSH effect can be realized in newly found intrinsic AFM TI materials (MnBi$_2$Te$_4$)$_m$(Bi$_2$Te$_3$)$_n$ or magnetic-doped TI multilayers by current experimental setups.

\end{abstract}

\maketitle

{\emph{Introduction.}}---Searching for topological insulators (TIs) has been one of the focus areas in condensed matter physics in the past years~\cite{KaneRMP2010,ZhangRMP2011}.
 Among them, a quantum spin-Hall (QSH) insulator which supports helical edge modes, is a two-dimensional (2D) $Z_2$ TI and can only survive in the presence of time-reversal symmetry (TRS) ~\cite{KanePRL2005,KanePRL05,Andrei2006,Konig2007,Roth2009,Du2011,Qian2014,Song2018,Wang2016}.
 In reality, such helical edge modes are generally subjected to inelastic backscattering even without TRS-breaking, which would ruin the ballistic transport mediated by helical edge modes~\cite{Konig2007,Jiang2009,Ivan2014}.
Therefore, a truly quantized helical-edge conductance was only experimentally observed in mesoscopic samples within several micrometers.
 On the other hand,  the QAH state that breaks TRS is characterized by the Chern number~(CN) and can support quantized conductance carried by chiral edge states in millimetre-size samples \cite{Haldane1988,Yurui2010,Chang2013,Liu2016,Heke2018,Mogi2017,Mogi2017b,Chang2018}. 
Moreover, higher-order topological insulators, which generalize the concept of TI, were recently proposed~\cite{Benalcazar2017,Langbehn2017,Benalcazar2017B,Song2017,Ezawa2018,Ezawa2018B,Yue2019,Baireuther2014}. The higher-order topology gives rise to one-dimensional (1D) helical and chiral states localized on the hinges of 3D second-order topological insulators (SOTIs)~\cite{Schindler2018,Schindler2018B,Liu2019}.

 Recently, there are tremendous progresses in fabricating magnetic TIs experimentally ~\cite{MBE2018,otrokov2018,Heke2019,Lieaaw5685,Vidal2018,hu2019van,wu2019,li2019dirac,hao2019gapless,chen2019topological}.
 Nobtaly, a breakthrough was made in synthesizing intrinsic 3D antiferromagnetic (AFM) TI material MnBi$_{2}$Te$_{4}$, which has van der Waals bonded layer structures. Soon after, both the QAH states and axion insulators were observed in MnBi$_{2}$Te$_{4}$ \cite{yuanbo2019,wangjian2019,Yayu2019,Wangjing}.
  Interestingly, (MnBi$_2$Te$_4$)$_m$(Bi$_2$Te$_3$)$_n$ family materials (such as, MnBi$_{4}$Te$_{7}$ and MnBi$_{6}$Te$_{10}$), constructed by stacking MnBi$_{2}$Te$_{4}$ septuplelayers and quintuple Bi$_{2}$Te$_{3}$ layers with different ratio $m/n$, are also realized in experiments \cite{vidal2019topological,Shi_2019,ding2019crystal,hu2019universal,xu2019persistent,tian2019magnetic}. Therefore, the (MnBi$_2$Te$_4$)$_m$(Bi$_2$Te$_3$)$_n$ materials are highly tunable and become a versatile platform to realized various topological phases.


 \begin{figure}[bht]
\centering
\includegraphics[width=3.2in]{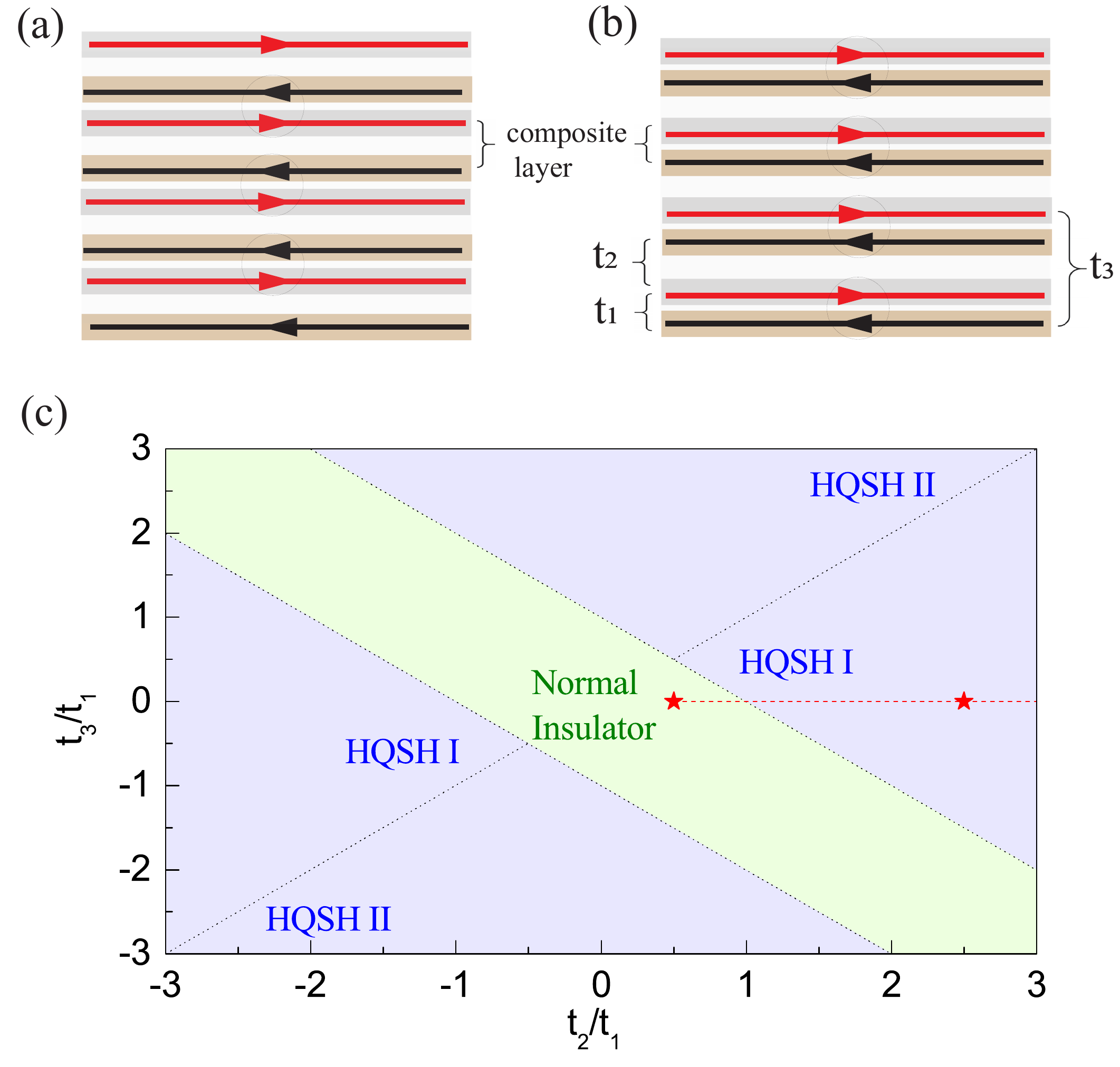}
\caption{(Color online). Schematic plots of alternatively stacked AFM TIs and the phase diagram. Two adjacent magnetic TI layers in opposite magnetization carrying chiral states with opposite charality~(marked by red and black arrows) form a composite layer. (a) Magnetic TIs between two adjacent composite layers are hybridized pairwise when the interlayer coupling $t_2$ dominates over the intralayer coupling $t_1$, leaving two chiral edge modes localized on the top and bottom layers. They are equivalent to a pair of helical edge modes in the QSH effect. On the contrary, (b) all the chiral edge modes within the composite layers hybridize in pairs if $t_1$ dominates. (c) Phase diagram of AFM TIs  on the plane of $t_2/t_1$ and $t_3/t_1$, with $t_{2,3}$ hopping between two adjacent composite layers. The two red stars on red line ($t_3=0$) indicate the parameter region studied in Fig.~\ref{fig2}.
\label{fig1} }
\end{figure}

 In this Letter, we propose a TRS-breaking QSH effect characterized by a pair of helical hinge modes (HHMs) in AFM TI multilayers. The HHMs consist of two spatially separated chiral hinge states with opposite chirality, which are localized on the hinges of the top and bottom surfaces of 3D AFM TIs. The QSH effect associated with the HHMs can be regarded as a topological characteristic of an SOTI in the 3D system, and we name this QSH effect as hinged quantum spin-Hall (HQSH) effect.
  To realize the HQSH effect, we design the AFM TI by alternatively stacking magnetic TIs (e.g. MnBi$_2$Te$_4$ septuplelayers) with opposite magnetization as shown in Fig.~\ref{fig1}. Two adjacent magnetic TI layers in opposite magnetization carrying opposite CNs form a composite layer~[see Fig.~\ref{fig1}(a)]. Here, the intralayer and interlayer couplings can be controlled by introducing spacer layers. When the interlayer coupling dominates over intralayer coupling, each pair of  neighboring counter-propagating chiral states from two adjacent composite layers tend to hybridize in pair and develop a gap in the bulk, while the two unhybridized chiral states on the hinges of the top and bottom surfaces remain gapless and form a pair of spatially separated HHMs~[Fig.~\ref{fig1}(a)]. On the contrary, if the intralayer coupling dominates, the two chiral edge states within individual composite layers tend to hybridize and thus
all the edge modes are gapped out in pairs [see Fig.~\ref{fig1}(b)], resulting in a trivial insulator.

In comparison with the conventional QSH effect, the HQSH effect shows two distinguishing features:
(i) It survives the breaking of TRS, while the conventional QSH effect requires the protection by TRS \cite{Andrei2006,KanePRL05,KanePRL2005}. This extends the concept of QSH effect to TRS-breaking (magnetic) materials, which were believed not to support the QSH effect.
(ii) Unlike the ordinary helical edge/hinge states in the conventional QSH effect or the previous SOTIs  \cite{Schindler2018,Schindler2018B,Liu2019,Du2015,Konig2007,Wu2018},
the HHMs in the HQSH effect support a quantized conductance in macroscopic samples because they are sustainable to dephasing and magnetic disorder.
Thus the HQSH effect, which is protected by spin Chern number, represents a different kind of topological phenomenon.

{\em Model Hamiltonian.}--- The model Hamiltonian $\mathcal{H}$ of 3D AFM TI multilayers can be obtained by alternatively stacking 2D QAH Hamiltonian with opposite CNs in the $z$ direction, which can be written as
\begin{eqnarray}
\mathcal{H}\!\!&=&\!\! \sum_{i=1}^{N_z}\Psi_{i}^{\dagger} (h_{D}\zeta_z \!+\! t_1\zeta_x) \Psi_{i} \!+\! [\Psi_{i}^{\dagger} (t_2 \zeta_{+} \!+\!t_3 \zeta_{-}) \Psi_{i+\!1}+ h.c.],\nonumber
\end{eqnarray}
where $\Psi_{n}^\dagger$ is electron creation operator in the $n$th composite layer, and $h_{D}= v_{f}(\sigma_y\sin k_{x}+\sigma_x\sin k_{y})+ (\Delta_z+m_k)\sigma_z$ is the QAH insulator Hamiltonian. The Pauli matrices  $\sigma_{x,y,z}$ and $\zeta_{x,y,z}$ act on the spin and layer spaces, respectively, and $\zeta_\pm=(\zeta_x\pm i\zeta_y)/2$.
The mass term $m_k=m_{0}-m_{1}(2-\cos k_{x}-\cos k_{y})$ with $m_0<0$ and $m_1>0$,
where $k_{x,y}$ is the wave vector in the $x$($y$) direction. $\Delta_z$ is the strength of exchange field and $v_{f}$ is the Fermi velocity. The Chern numbers of the QAH states are $C_\pm=\pm1$ for $\pm h_{D}$ when $\Delta_z>|m_0|$.
Note that we fix $\Delta=0.7$, $m_0=-0.1$, $m_1=1$ and $v_f=1$ in the following. Each two neighboring QAH with opposite CNs form a composite layer, and $i$ is the composite layer index. The QAH states are coupled by the intralayer hopping $t_{1}$ and interlayer hopping $t_{2,3}$ that are determined by the thickness of spacer layers [see Figs.~\ref{fig1}(a)-\ref{fig1}(b)].
At the beginning, we set $t_3 = 0$ and discuss how the topological phase of the system is determined by the intralayer and interlayer hoppings $t_1$ and $t_2$. In the limited case of $t_1\neq0$ and $t_2=0$, the AFM TIs are decoupled into $N$ isolated composite layers, and the two QAH insulators within a composite layer hybridize to gap out the two chiral states.
On the contrary, when $t_1=0$ and $t_2\neq0$, each two neighboring QAH insulators from two adjacent composite layers are coupled by $t_2$ and there are two isolated QAH insulators localized on the top and bottom composite layers, leaving a pair of HHMs in 3D AFM TIs. We name it the HQSH effect.

If we regard the 2D Hamiltonian $h_{D}$ as a model parameter, $\mathcal{H}$ is exactly a 3D generalization of the Rice-Mele model \cite{SSH1979,RM1982}, which consists of $2N_z$ quasi-2D bands related to $h_{D}$.
When $|t_1|<|t_2|$, there exist two quasi-2D bands localized on the top and bottom surfaces, originating from the two end states of the Rice-Mele model \cite{SSH1979,RM1982}. Then we can obtain an effective Hamiltonian by performing a projection $H_{proj}=\bigoplus \hat{P}_{\pm} \mathcal{H} \hat{P}_{\pm}=\bigoplus\pm h_{D}\hat{P}_{\pm}$ at the large $N_z$ limit, where the projector operator $\hat{P}_{\pm}=|\nu_{0z}^{\pm}\rangle\langle\nu_{0z}^{\pm}|$. The wave functions $|\nu_{0z}^{+}\rangle=A_+\lambda^{z}[1,0]^T$ and $|\phi_z^-\rangle=A_-\lambda^{-z}[0,1]^T$ with $\lambda =t_1/t_2$ and normalization constants $A_\pm$. Therefore, the whole system can be characterized by
a quantized spin CN $C_s\equiv(C_+ - C_-)/2=1$, where the CNs $C_{\pm}$ of two QAH subsystems can be calculated by the non-commutative Kubo formula $C_{\pm} = 2\pi i \Tr[Q_{\pm}[\partial_{k_x} Q_{\pm},\partial_{k_y}Q_{\pm}]]$
with $Q_{\pm}=QP_{\pm}$ and the projector $Q$ onto the occupied states of $\mathcal{H}$ \cite{Prodan2009,Prodan2011,Sheng2005,Sheng2006}.

\begin{figure}[tbh]
	\centering
	\includegraphics[width=3in]{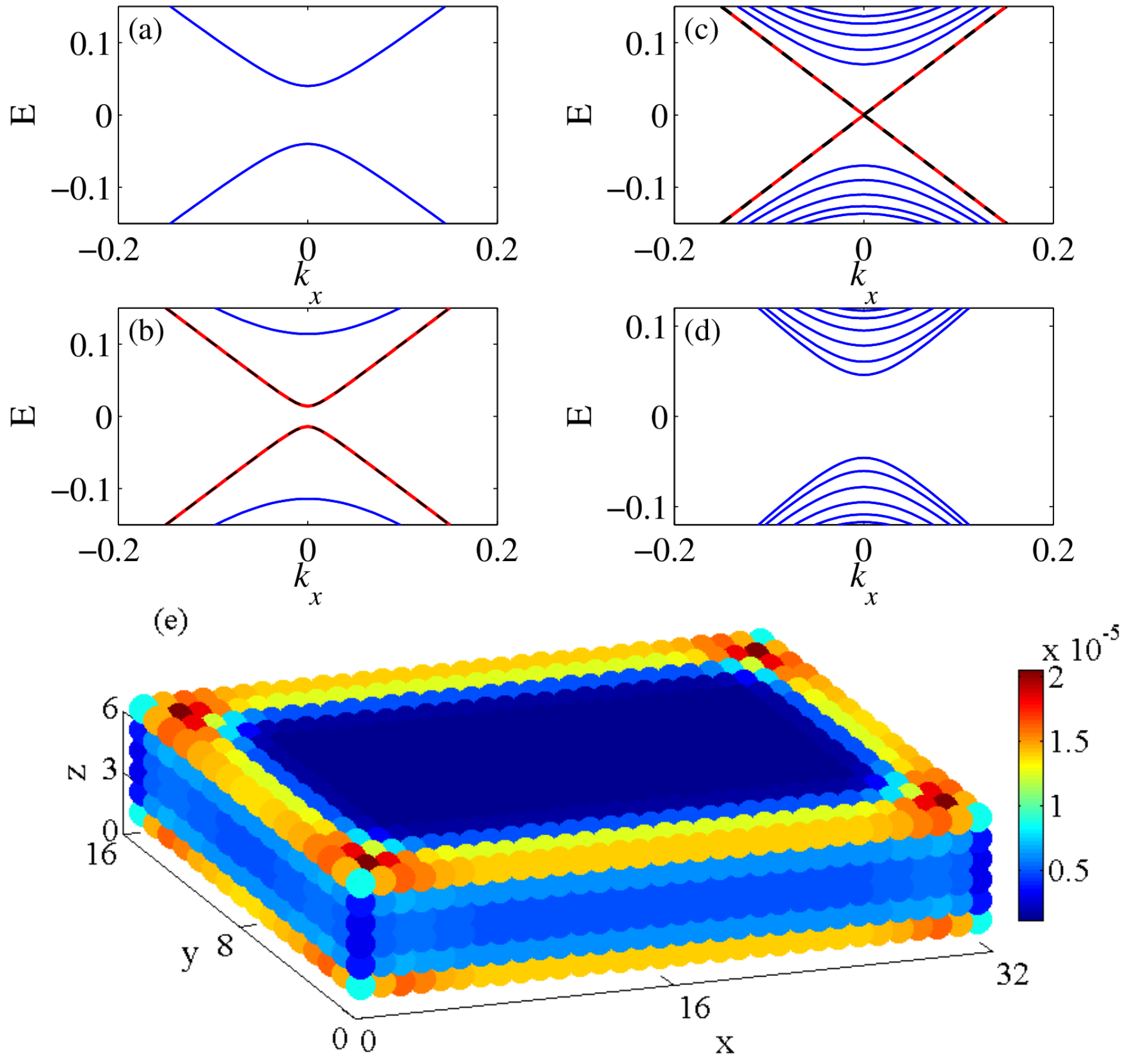}
	\caption{(Color online). The  energy  band  structure  of  AFM TI multilayers with the width $N_y=16$, thickness $N_z$ and intralayer coupling $t_1$ and interlayer coupling $t_2$.
		(a) The two chiral edge modes are hybridized and gapped out for $N_z=1$.
		(b)-(c) For $t_1=0.04<t_2=0.1$, the chiral edge modes in different composite layers are hybridized and gapped out in pairs, leaving two isolated chiral edge modes [see the dashed red and black lines in (b) $N_z=2$ and (c) $N_z=6$] localized on the top and bottom layers. (e) Local density of state with $N_x\times N_y\times N_z=32\times 16\times 6$ at $E =0.01$. (d) For $t_1=0.08>t_2=0.04$, all the chiral edge modes with opposite chirality in same composite layers are coupled and gapped out in pairs with $N_z=6$.
		\label{fig2} }
\end{figure}
In virtue of the bulk-boundary correspondence, the effective Hamiltonian of the HHMs can be obtained by a projection similarly. The effective Hamiltonian for two chiral edge states of the neighboring QAH insulators in one composite layer can be written as $H_{1D}= v_f k_x\zeta_z$. Then the surface states of the AFM TIs on the $x-z$ plane can be expressed by $H_{1D}$ as $H_{surf}=\sum_{n=1}^{N_z}\psi_{n}^{\dagger} (v_f k_x\zeta_z + t_1\zeta_x) \psi_{n} + \psi_{n}^{\dagger} t_2 \zeta_{+} \psi_{n+1} + h.c.$ with the energy spectrum $E_k=\pm(v_f^2 k_x^2 + (t_1 + t_2\cos k_z)^2+t_2^2\sin^2 k_z)^{1/2}$. The gap at $k_x=0$ closes when $|t_1|=|t_2|$, giving rise to a topologically nontrivial phase with $|t_1|<|t_2|$ and a trivial phase with $|t_1|>|t_2|$. In fact, at $k_x=0$, $H_{surf}$ is exactly reduced to the Su-Schrieffer-Heeger~(SSH) model which has two zero-energy states localized on the boundaries for $|t_1|<|t_2|$, with the wave functions $|\nu_{0z}^{\pm}\rangle$ at the large $N_z$ limit \cite{SSH1979,RM1982}. Correspondingly, the HHMs on the $xz$ plane can be captured by the effective Hamiltonian $\langle\nu_{0z}^\pm| H_{surf} |\nu_{0z}^\pm\rangle=\pm\hbar v_f k_x$.

If we now turn on $t_3$, the phase diagram [Fig.~\ref{fig1}(c)] of the AFM TIs can be determined by the gap closing points of the surface state spectrum $E_k=\pm\{v_f^2 k_x^2 + [t_1 + (t_2+t_3)\cos k_z]^2+(t_2-t_3)^2\sin^2 k_z\}^{1/2}$.
Generally, the HQSH and normal insulator phases are separated by $|t_1|=|t_2+t_3|$, furthermore the HQSH are divided into two parts (HQSH I and HQSH II) by $t_2=t_3$. Note that the HHMs in the HQSH I and HQSH II has opposite chirality.
That's because the two QAH insulators and thus the chiral edge states in the composite layer switch positions if we exchange $t_2$ and $t_3$ in the Hamiltonian $\mathcal{H}$.
In the following, we focus on the HQSH I in the phase diagram and fix $t_3=0$ unless otherwise specified.

 To demonstrate the emergence of the HHMs in the AFM TI multilayers, we numerically investigate the energy band structure of the Hamiltonian $\mathcal{H}$ as varying the composite layer number $N_z$ in Fig.~\ref{fig2}. In general, the energy spectra are two-fold degenerate, because $\mathcal{H}$ respects a combined symmetry of TRS and inversion symmetry.
Figure~\ref{fig2}(a) shows two chiral edge modes with opposite chirality are coupled and gapped out when $N_z=1$.
For $t_2=0.1>t_1=0.04$, when $N_z>1$, the chiral edge modes from two adjacent composite layers tend to hybridized in pairs, leaving two sets of isolated chiral edge modes [see the dashed red and black lines in Fig.~\ref{fig2}(b) $N_z = 2$ and Fig. \ref{fig2}(c) $N_z = 6$] localized on the top and bottom layers. Note that the system has a small hybridization gap for $N_z=2$ due to the finite-size effect, which is closed for a larger size $N_z=6$. Furthermore, under the open boundary condition along all the three directions, we show the local density of states (DOS) in Fig.~\ref{fig2}(e) at $E=0.01$ and find that the midgap states are localized on the hinges of the sample surfaces. On the contrary, fixing $N_z=6$, when $t_2=0.04<t_1=0.08$, the chiral edge modes in the same composite layers are hybridized in pairs, leaving an energy gap in the surface band spectrum as shown in Fig.~\ref{fig2}(d).
Therefore, the HQSH characterized by a pair of HHMs is realized in AFM TIs when the interlayer coupling $t_2$ dominates over the intralayer one. 

To explore robustness of the HQSH effect, we study the conductance in the presence of TRS-breaking disorder by using the Landauer-Buttiker formula. The current in the lead $p$ can be expressed as:
$I_p=e^2/h\sum_{q\neq p}T_{pq}(E_F)(V_p-V_q)$ where $V_p$ is the bias in the lead $p$
and $T_{pq}(E_F)= \Tr[\Gamma_p G^r \Gamma_q G^a ] $ is the transmission coefficient from the lead $q$ to $p$ with the Green's function $G^r=[G^a]^\dagger$ and the line width function $\Gamma_p $ \cite{datta1995}.
Here, TRS-breaking Anderson disorder is included as a random on-site potential $V(r)$ on each orbital independently, which is uniformly distributed in the range of $[-W/2,W/2]$.
In the following, we systematically evaluate the longitudinal conductance $G_{12,12}$ and the nonlocal conductance $G_{14,23}$ for the two-terminal device and the $\pi$ device, respectively \cite{Jiang2014,Du2015}.

\begin{figure}[bht]
\centering
\includegraphics[width=3.2in]{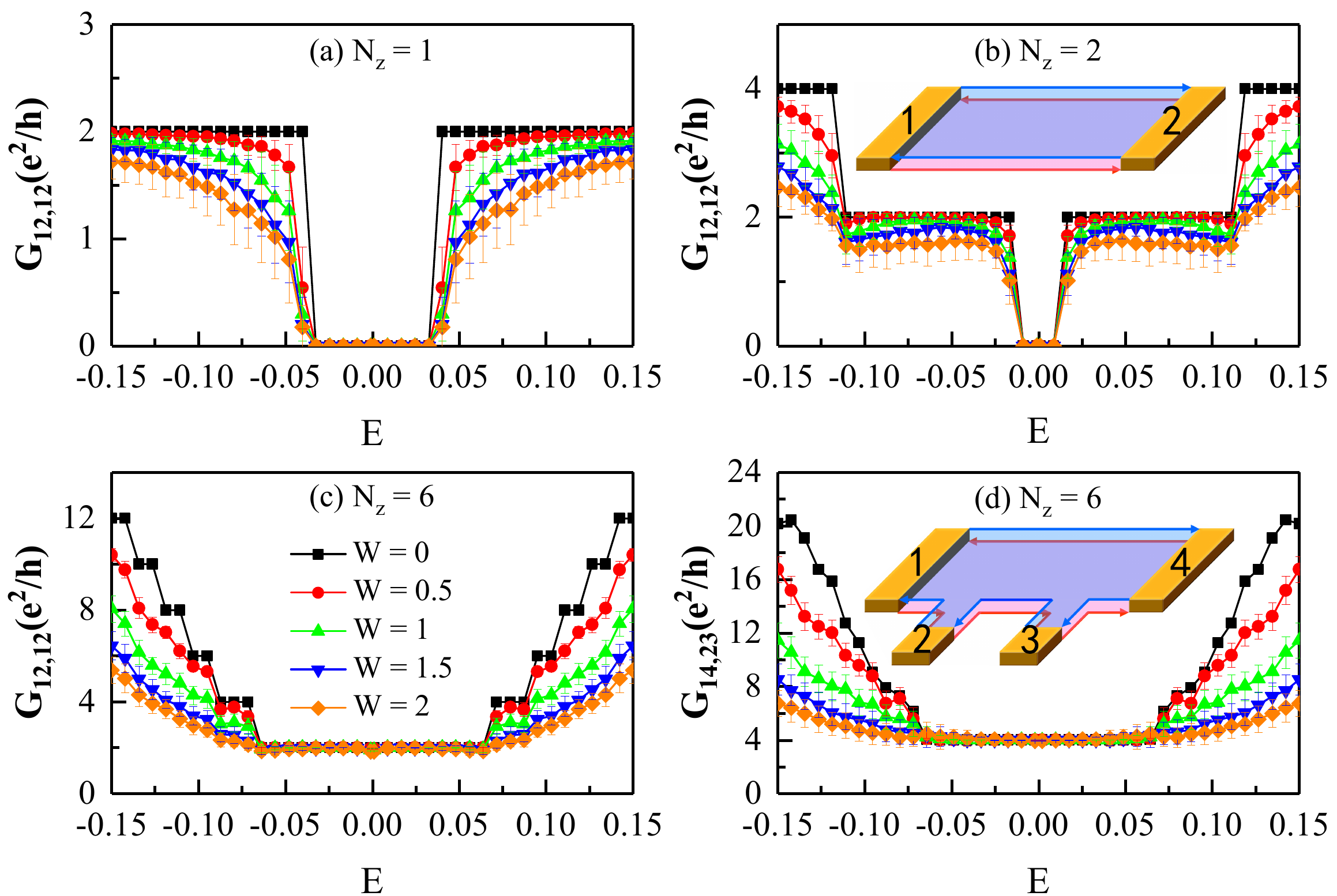}
\caption{(Color online).
 Plots of  (a-c) the two-terminal conductance $G_{12,12}$  and (d) nonlocal conductance $G_{14,23}$ versus the Fermi energy $E$ for different thickness (a) $N_z=1$, (b) $2$, and [(c),(d)] $6$ at various disorder strength $W$. The errorbar denotes conductance fluctuation. The three sets of model parameters are the same as those in Figs.~\ref{fig2}(a)-\ref{fig2}(c), respectively, with the sample size $N_x\times N_y\times N_z=64\times16\times N_z$ in Figs.~\ref{fig3}(a)-\ref{fig3}(c) and $96\times16\times6$ in Fig.~\ref{fig3}(d). The insets of (b) and (d) show a two-terminal device and a $\pi$-bar device, respectively.
\label{fig3} }
\end{figure}
In Figs.~\ref{fig3}(a)-\ref{fig3}(b), we calculate the longitudinal conductance $G_{12,12}$ in the two-terminal device [see the inset of Fig.~\ref{fig3}(b)] as a function of the Fermi energy $E$ for different thickness $N_z$ at various disorder strengths $W$. 
 We set the system in the topologically nontrivial region with $t_1=0.04<t_2=0.1$. For $N_z=1$ in Fig.~\ref{fig3}(a), we find an insulating gap with $G_{12,12}=0$ between a quantized conductance plateau $G_{12,12}=2e^2/h$ at $W=0$, which loses quantization in the presence of disorder. That's because the two chiral edge modes are coupled and thus not topologically protected as shown Fig.~\ref{fig2}(a). In Fig.~\ref{fig3}(b), there are four chiral edge modes with $G_{12,12}=4e^2/h$ for $N_z=2$ and the finite-size effect gap with $G_{12,12}=0$ decreases.

  By increasing the thickness to $N_z=6$ in Fig.~\ref{fig3}(c), it is found that a quantized conductance plateau $G_{12,12}=2e^2/h$ shows up near the band center $E=0$, because the finite-size gap becomes negligible [see Fig.~\ref{fig2}(c)]. Notably, the plateau $G_{12,12}=2e^2/h$ is topologically protected and remains stable under disorder with strength larger than the band gap, while the higher-order plateaus $2\nu e^2/h$ ($\nu=2,3,4,...$) lose quantization in the presence of disorder. Therefore, we conclude that the HHMs in the HQSH are robust to TRS-breaking disorder, distinct from helical edge modes in the conventional QSH effect.
At last, we propose the HHMs can be detected by a nonlocal conductance $G_{14,23}$ in a $\pi$ device consisting of a sample attached to four leads [inset of Fig.~\ref{fig3}(d)]. Notably, the first conductance plateau $G_{14,23}=4e^2/h$ that indicates the existence of the HHMs, is robust to disorder.

\begin{figure}[bht]
	\centering
	\includegraphics[width=3.2in]{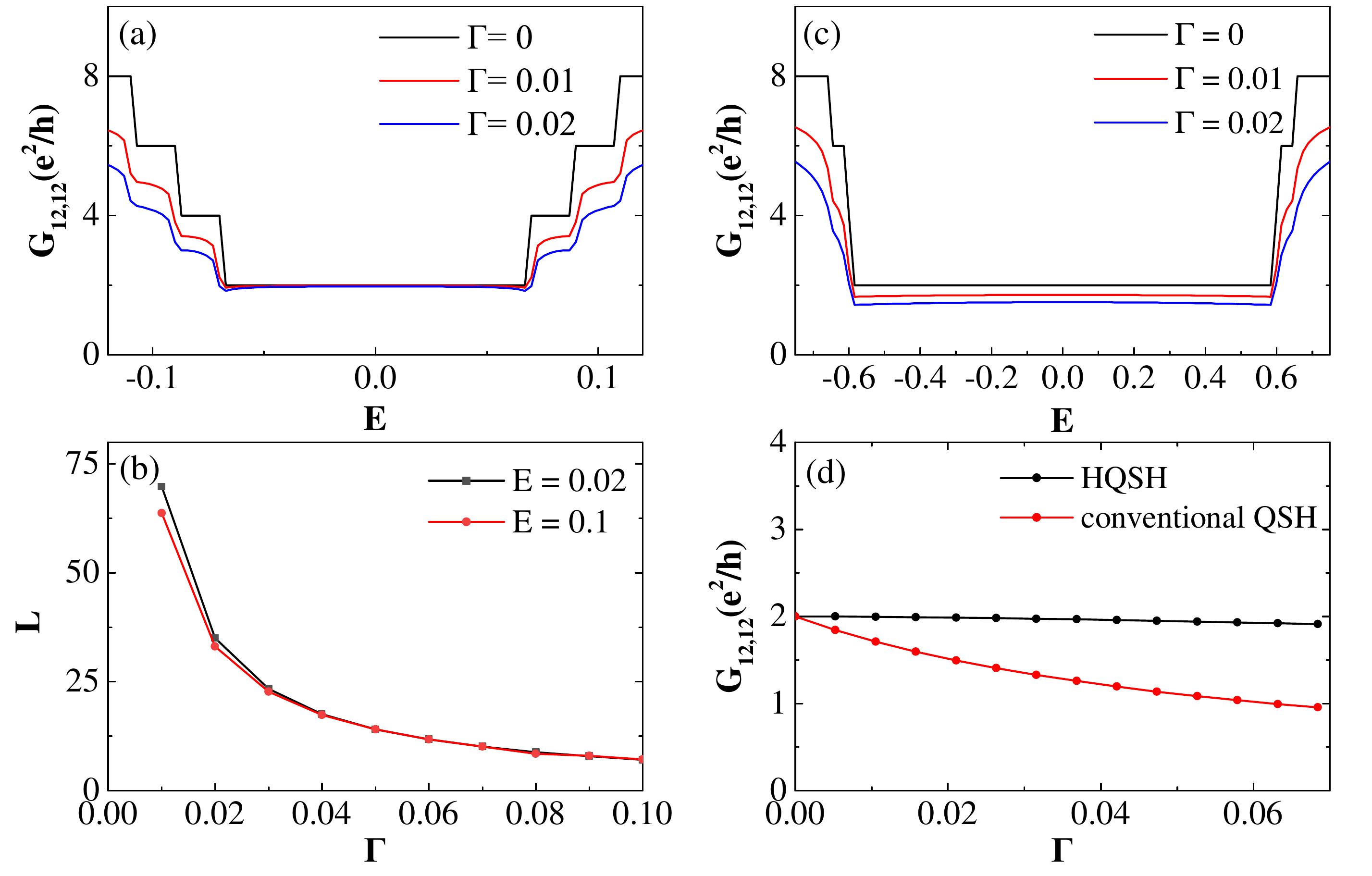}
	\caption{(Color online). (a)Two-terminal conductance $G_{12,12}$ of the HQSH versus the Fermi energy $E$ for various dephasing strength $\Gamma$.  (b) Phase coherent length $L_{\phi}$ as a function of $\Gamma$.
(c) $G_{12,12}$ of the conventional QSH in the Bernevig-Hughes-Zhang (BHZ) model versus $E$ for various $\Gamma$.
(d) $G_{12,12}$ versus $\Gamma$ for the HQSH and the conventional QSH at $E=0.02$.
The parameters of the Dirac Hamiltonian in the BHZ model are the same of the $h_D$ in the HQSH model.
The sample size is $N_x\times N_y\times N_z=32\times16\times6$.
\label{fig4} }
\end{figure}
Now let us come to investigate the inelastic scattering in the system by simulating
the dephasing effects on the HQSH~\cite{BM1986,Shi2001}.
In a realistic sample, there are plenty of possible dephasing processes because of electron-electron or electron-phonon interactions.
We introduce one virtual lead with the linewidth $\Gamma$ to each site to simulate the dephasing effect using Landauer-Buttiker formula~\cite{BM1986,Shi2001}.
$\Gamma$ is dephasing strength and directly related to the phase coherence length $L_\phi$~\cite{Xing2008,Jiang2009}, which is an experimentally measurable parameter.
Figure \ref{fig4}(b) shows the relation of the phase coherent length $L_\phi$ versus the dephasing strength $\Gamma$. With increasing $\Gamma$, $L_\phi$ decreases rapidly and monotonically.

Next, we present our numerical results of  the dephasing effect on the two-terminal conductance $G_{12,12}$. Figure \ref{fig4}(a) plots $G_{12,12}$ versus the Fermi energy $E$ for various $\Gamma$. It is clear that the quantized plateau at $G_{12,12}=2e^2/h$ due to the HHMs in the HQSH is hardly affected by the dephasing.
By contrast, we simulate the conventional QSH using the Bernevig-Hughes-Zhang model \cite{Andrei2006} in Fig.~\ref{fig4}(c) and find the conductance plateau at $G_{12,12}=2e^2/h$ loses quantization dramatically in the presence of the (spin) dephasing~\cite{Konig2007,Jiang2009,Ivan2014}.
To be specifically, as shown in Fig.~\ref{fig4}(d), we find that the quantized conductance $G_{12,12}=2e^2/h$ at $E=0.02$ decreases by less than $5\%$ at dephasing strength $\Gamma=0.07$, even when the system size $L=32$ far exceeds the phase coherent length $L_\phi\approx10$.
That's because the HHMs in HQSH are spatially separated as discussed above, which thus behave as the chiral edge channel in the quantum Hall effect that is hardly affected the dephasing.
In contrast, the conductance plateau $G_{12,12}=2e^2/h$ in the conventional QSH at $E=0.02$ decreases by $7\%$ even for $\Gamma=0.005$ and more than $52\%$ for $\Gamma=0.07$ [see Fig.~\ref{fig4}(d)].
Therefore, we conclude that the HHMs in the HQSH is robust to the dephasing (inelastic scattering), which can be observed in macroscopic samples.

Experimental Materials.-- The proposed HQSH effect in 3D AFM TI multilayers can be readily realized in realistic materials. Recently, the intrinsic AFM TIs family materials (MnBi$_2$Te$_4$)$_m$(Bi$_2$Te$_3$)$_n$ are fabricated experimentally  \cite{vidal2019topological,Shi_2019,ding2019crystal,hu2019universal,xu2019persistent,tian2019magnetic}, which can be regarded as stacked  MnBi$_{2}$Te$_{4}$ septuplelayers (denoted as ``A") and Bi$_{2}$Te$_{3}$ quintuple (denoted as ``B")
layers. If we introduce Bi$_{2}$Te$_{3}$ quintuple layers to intrinsic 3D AFM TI MnBi$_{2}$Te$_{4}$, the new system forms a superlattice structure with different interlayer and intralayer couplings. For example, in ``ABBA..." superlattices, the interlayer coupling (between two neighbouring MnBi$_{2}$Te$_{4}$ septuplelayers) in ``AA" is dominant over the intralayer coupling in ``ABBA" due to $B$ layers.
Moreover, the magnetically doped (Bi,Sb)$_2$Te$_3$ TI and CdSe normal insulator layers in the superlattice structure has been successfully grown by molecular beam epitaxy \cite{MBE2018}. It has been shown that the coercive field of magnetic TI in each layer can be controlled by mixing the dopants Cr and V at varied ratios~\cite{Yunbo2018}. In fact, a magnetic TI bilayer with different dopants ratios was driven to the opposite magnetization by sweeping the magnetic field in a recent experiment~\cite{Fengyang2019}.
Therefore, it is natural to expect that HQSH effect in 3D AFM TIs can be realized in the (MnBi$_2$Te$_4$)$_m$(Bi$_2$Te$_3$)$_n$ family materials as well as magnetic doped TIs multilayers.

Summary.-- In conclusion, we propose to realize the HQSH effect with a pair of HHMs in 3D AFM TIs, which can be fabricated by current experimental setups. These HHMs can be identified by a nonlocal conductance plateau of $4e^2/h$ in $\pi$-bar measurement. In contrast to the conventional QSH effect, the HQSH effect has a very different topological origin and thus can survive without TRS, where the HHMs therefore is robust to magnetic impurities as well as spin dephasing.



{\emph{Acknowledgement.}}---  We thank Hua Jiang and Haiwen Liu for illuminating discussions.  This work is financially supported by NBRPC (Grants No. 2015CB921102), NSFC (Grants No. 11534001, No. 11822407, No. 11704106), and supported
by the Fundamental Research Funds for the Central Universities.
C.-Z. Chen are also funded by the Priority Academic Program Development of Jiangsu Higher Education Institutions. D.-H.X. is also supported by the Chutian Scholars Program in Hubei Province.

{\emph{Note added.}}--- Recently, we became aware of a complementary study \cite{EK2019}, which treats a similar situation but focuses on a different aspect.


\begin{thebibliography}{68}%
\makeatletter
\providecommand \@ifxundefined [1]{%
 \@ifx{#1\undefined}
}%
\providecommand \@ifnum [1]{%
 \ifnum #1\expandafter \@firstoftwo
 \else \expandafter \@secondoftwo
 \fi
}%
\providecommand \@ifx [1]{%
 \ifx #1\expandafter \@firstoftwo
 \else \expandafter \@secondoftwo
 \fi
}%
\providecommand \natexlab [1]{#1}%
\providecommand \enquote  [1]{``#1''}%
\providecommand \bibnamefont  [1]{#1}%
\providecommand \bibfnamefont [1]{#1}%
\providecommand \citenamefont [1]{#1}%
\providecommand \href@noop [0]{\@secondoftwo}%
\providecommand \href [0]{\begingroup \@sanitize@url \@href}%
\providecommand \@href[1]{\@@startlink{#1}\@@href}%
\providecommand \@@href[1]{\endgroup#1\@@endlink}%
\providecommand \@sanitize@url [0]{\catcode `\\12\catcode `\$12\catcode
  `\&12\catcode `\#12\catcode `\^12\catcode `\_12\catcode `\%12\relax}%
\providecommand \@@startlink[1]{}%
\providecommand \@@endlink[0]{}%
\providecommand \url  [0]{\begingroup\@sanitize@url \@url }%
\providecommand \@url [1]{\endgroup\@href {#1}{\urlprefix }}%
\providecommand \urlprefix  [0]{URL }%
\providecommand \Eprint [0]{\href }%
\providecommand \doibase [0]{http://dx.doi.org/}%
\providecommand \selectlanguage [0]{\@gobble}%
\providecommand \bibinfo  [0]{\@secondoftwo}%
\providecommand \bibfield  [0]{\@secondoftwo}%
\providecommand \translation [1]{[#1]}%
\providecommand \BibitemOpen [0]{}%
\providecommand \bibitemStop [0]{}%
\providecommand \bibitemNoStop [0]{.\EOS\space}%
\providecommand \EOS [0]{\spacefactor3000\relax}%
\providecommand \BibitemShut  [1]{\csname bibitem#1\endcsname}%
\let\auto@bib@innerbib\@empty
\bibitem [{\citenamefont {Hasan}\ and\ \citenamefont
  {Kane}(2010)}]{KaneRMP2010}%
  \BibitemOpen
  \bibfield  {author} {\bibinfo {author} {\bibfnamefont {M.~Z.}\ \bibnamefont
  {Hasan}}\ and\ \bibinfo {author} {\bibfnamefont {C.~L.}\ \bibnamefont
  {Kane}},\ }\href {\doibase 10.1103/RevModPhys.82.3045} {\bibfield  {journal}
  {\bibinfo  {journal} {Rev. Mod. Phys.}\ }\textbf {\bibinfo {volume} {82}},\
  \bibinfo {pages} {3045} (\bibinfo {year} {2010})}\BibitemShut {NoStop}%
\bibitem [{\citenamefont {Qi}\ and\ \citenamefont
  {Zhang}(2011)}]{ZhangRMP2011}%
  \BibitemOpen
  \bibfield  {author} {\bibinfo {author} {\bibfnamefont {X.-L.}\ \bibnamefont
  {Qi}}\ and\ \bibinfo {author} {\bibfnamefont {S.-C.}\ \bibnamefont {Zhang}},\
  }\href {\doibase 10.1103/RevModPhys.83.1057} {\bibfield  {journal} {\bibinfo
  {journal} {Rev. Mod. Phys.}\ }\textbf {\bibinfo {volume} {83}},\ \bibinfo
  {pages} {1057} (\bibinfo {year} {2011})}\BibitemShut {NoStop}%
\bibitem [{\citenamefont {Kane}\ and\ \citenamefont
  {Mele}(2005{\natexlab{a}})}]{KanePRL2005}%
  \BibitemOpen
  \bibfield  {author} {\bibinfo {author} {\bibfnamefont {C.~L.}\ \bibnamefont
  {Kane}}\ and\ \bibinfo {author} {\bibfnamefont {E.~J.}\ \bibnamefont
  {Mele}},\ }\href {\doibase 10.1103/PhysRevLett.95.226801} {\bibfield
  {journal} {\bibinfo  {journal} {Phys. Rev. Lett.}\ }\textbf {\bibinfo
  {volume} {95}},\ \bibinfo {pages} {226801} (\bibinfo {year}
  {2005}{\natexlab{a}})}\BibitemShut {NoStop}%
\bibitem [{\citenamefont {Kane}\ and\ \citenamefont
  {Mele}(2005{\natexlab{b}})}]{KanePRL05}%
  \BibitemOpen
  \bibfield  {author} {\bibinfo {author} {\bibfnamefont {C.~L.}\ \bibnamefont
  {Kane}}\ and\ \bibinfo {author} {\bibfnamefont {E.~J.}\ \bibnamefont
  {Mele}},\ }\href {\doibase 10.1103/PhysRevLett.95.146802} {\bibfield
  {journal} {\bibinfo  {journal} {Phys. Rev. Lett.}\ }\textbf {\bibinfo
  {volume} {95}},\ \bibinfo {pages} {146802} (\bibinfo {year}
  {2005}{\natexlab{b}})}\BibitemShut {NoStop}%
\bibitem [{\citenamefont {Bernevig}\ \emph {et~al.}(2006)\citenamefont
  {Bernevig}, \citenamefont {Hughes},\ and\ \citenamefont
  {Zhang}}]{Andrei2006}%
  \BibitemOpen
  \bibfield  {author} {\bibinfo {author} {\bibfnamefont {B.~A.}\ \bibnamefont
  {Bernevig}}, \bibinfo {author} {\bibfnamefont {T.~L.}\ \bibnamefont
  {Hughes}}, \ and\ \bibinfo {author} {\bibfnamefont {S.-C.}\ \bibnamefont
  {Zhang}},\ }\href {\doibase 10.1126/science.1133734} {\ \textbf {\bibinfo
  {volume} {314}},\ \bibinfo {pages} {1757} (\bibinfo {year}
  {2006})}\BibitemShut {NoStop}%
\bibitem [{\citenamefont {K{\"o}nig}\ \emph {et~al.}(2007)\citenamefont
  {K{\"o}nig}, \citenamefont {Wiedmann}, \citenamefont {Br{\"u}ne},
  \citenamefont {Roth}, \citenamefont {Buhmann}, \citenamefont {Molenkamp},
  \citenamefont {Qi},\ and\ \citenamefont {Zhang}}]{Konig2007}%
  \BibitemOpen
  \bibfield  {author} {\bibinfo {author} {\bibfnamefont {M.}~\bibnamefont
  {K{\"o}nig}}, \bibinfo {author} {\bibfnamefont {S.}~\bibnamefont {Wiedmann}},
  \bibinfo {author} {\bibfnamefont {C.}~\bibnamefont {Br{\"u}ne}}, \bibinfo
  {author} {\bibfnamefont {A.}~\bibnamefont {Roth}}, \bibinfo {author}
  {\bibfnamefont {H.}~\bibnamefont {Buhmann}}, \bibinfo {author} {\bibfnamefont
  {L.~W.}\ \bibnamefont {Molenkamp}}, \bibinfo {author} {\bibfnamefont {X.-L.}\
  \bibnamefont {Qi}}, \ and\ \bibinfo {author} {\bibfnamefont {S.-C.}\
  \bibnamefont {Zhang}},\ }\href {\doibase 10.1126/science.1148047} {\ \textbf
  {\bibinfo {volume} {318}},\ \bibinfo {pages} {766} (\bibinfo {year}
  {2007})}\BibitemShut {NoStop}%
\bibitem [{\citenamefont {Roth}\ \emph {et~al.}(2009)\citenamefont {Roth},
  \citenamefont {Br{\"u}ne}, \citenamefont {Buhmann}, \citenamefont
  {Molenkamp}, \citenamefont {Maciejko}, \citenamefont {Qi},\ and\
  \citenamefont {Zhang}}]{Roth2009}%
  \BibitemOpen
  \bibfield  {author} {\bibinfo {author} {\bibfnamefont {A.}~\bibnamefont
  {Roth}}, \bibinfo {author} {\bibfnamefont {C.}~\bibnamefont {Br{\"u}ne}},
  \bibinfo {author} {\bibfnamefont {H.}~\bibnamefont {Buhmann}}, \bibinfo
  {author} {\bibfnamefont {L.~W.}\ \bibnamefont {Molenkamp}}, \bibinfo {author}
  {\bibfnamefont {J.}~\bibnamefont {Maciejko}}, \bibinfo {author}
  {\bibfnamefont {X.-L.}\ \bibnamefont {Qi}}, \ and\ \bibinfo {author}
  {\bibfnamefont {S.-C.}\ \bibnamefont {Zhang}},\ }\href {\doibase
  10.1126/science.1174736} {\bibfield  {journal} {\bibinfo  {journal}
  {Science}\ }\textbf {\bibinfo {volume} {325}},\ \bibinfo {pages} {294}
  (\bibinfo {year} {2009})}\BibitemShut {NoStop}%
\bibitem [{\citenamefont {Knez}\ \emph {et~al.}(2011)\citenamefont {Knez},
  \citenamefont {Du},\ and\ \citenamefont {Sullivan}}]{Du2011}%
  \BibitemOpen
  \bibfield  {author} {\bibinfo {author} {\bibfnamefont {I.}~\bibnamefont
  {Knez}}, \bibinfo {author} {\bibfnamefont {R.-R.}\ \bibnamefont {Du}}, \ and\
  \bibinfo {author} {\bibfnamefont {G.}~\bibnamefont {Sullivan}},\ }\href
  {\doibase 10.1103/PhysRevLett.107.136603} {\bibfield  {journal} {\bibinfo
  {journal} {Phys. Rev. Lett.}\ }\textbf {\bibinfo {volume} {107}},\ \bibinfo
  {pages} {136603} (\bibinfo {year} {2011})}\BibitemShut {NoStop}%
\bibitem [{\citenamefont {Qian}\ \emph {et~al.}(2014)\citenamefont {Qian},
  \citenamefont {Liu}, \citenamefont {Fu},\ and\ \citenamefont
  {Li}}]{Qian2014}%
  \BibitemOpen
  \bibfield  {author} {\bibinfo {author} {\bibfnamefont {X.}~\bibnamefont
  {Qian}}, \bibinfo {author} {\bibfnamefont {J.}~\bibnamefont {Liu}}, \bibinfo
  {author} {\bibfnamefont {L.}~\bibnamefont {Fu}}, \ and\ \bibinfo {author}
  {\bibfnamefont {J.}~\bibnamefont {Li}},\ }\href {\doibase
  10.1126/science.1256815} {\bibfield  {journal} {\bibinfo  {journal}
  {Science}\ }\textbf {\bibinfo {volume} {346}},\ \bibinfo {pages} {1344}
  (\bibinfo {year} {2014})}\BibitemShut {NoStop}%
\bibitem [{\citenamefont {Song}\ \emph {et~al.}(2018)\citenamefont {Song},
  \citenamefont {Jia}, \citenamefont {Zhang}, \citenamefont {Zhu},
  \citenamefont {Shi}, \citenamefont {Wang}, \citenamefont {Zhu}, \citenamefont
  {Yuan}, \citenamefont {Zhang}, \citenamefont {Xing},\ and\ \citenamefont
  {Li}}]{Song2018}%
  \BibitemOpen
  \bibfield  {author} {\bibinfo {author} {\bibfnamefont {Y.-H.}\ \bibnamefont
  {Song}}, \bibinfo {author} {\bibfnamefont {Z.-Y.}\ \bibnamefont {Jia}},
  \bibinfo {author} {\bibfnamefont {D.}~\bibnamefont {Zhang}}, \bibinfo
  {author} {\bibfnamefont {X.-Y.}\ \bibnamefont {Zhu}}, \bibinfo {author}
  {\bibfnamefont {Z.-Q.}\ \bibnamefont {Shi}}, \bibinfo {author} {\bibfnamefont
  {H.}~\bibnamefont {Wang}}, \bibinfo {author} {\bibfnamefont {L.}~\bibnamefont
  {Zhu}}, \bibinfo {author} {\bibfnamefont {Q.-Q.}\ \bibnamefont {Yuan}},
  \bibinfo {author} {\bibfnamefont {H.}~\bibnamefont {Zhang}}, \bibinfo
  {author} {\bibfnamefont {D.-Y.}\ \bibnamefont {Xing}}, \ and\ \bibinfo
  {author} {\bibfnamefont {S.-C.}\ \bibnamefont {Li}},\ }\href {\doibase
  10.1038/s41467-018-06635-x} {\bibfield  {journal} {\bibinfo  {journal}
  {Nature Communications}\ }\textbf {\bibinfo {volume} {9}},\ \bibinfo {pages}
  {4071} (\bibinfo {year} {2018})}\BibitemShut {NoStop}%
\bibitem [{\citenamefont {Wang}\ \emph {et~al.}(2016)\citenamefont {Wang},
  \citenamefont {Zhang}, \citenamefont {Liu}, \citenamefont {Liu},
  \citenamefont {Tang}, \citenamefont {Song}, \citenamefont {Zhong},
  \citenamefont {Peng}, \citenamefont {Li}, \citenamefont {Nie}, \citenamefont
  {Wang}, \citenamefont {Zhou}, \citenamefont {Ma}, \citenamefont {Xue},\ and\
  \citenamefont {Liu}}]{Wang2016}%
  \BibitemOpen
  \bibfield  {author} {\bibinfo {author} {\bibfnamefont {Z.~F.}\ \bibnamefont
  {Wang}}, \bibinfo {author} {\bibfnamefont {H.}~\bibnamefont {Zhang}},
  \bibinfo {author} {\bibfnamefont {D.}~\bibnamefont {Liu}}, \bibinfo {author}
  {\bibfnamefont {C.}~\bibnamefont {Liu}}, \bibinfo {author} {\bibfnamefont
  {C.}~\bibnamefont {Tang}}, \bibinfo {author} {\bibfnamefont {C.}~\bibnamefont
  {Song}}, \bibinfo {author} {\bibfnamefont {Y.}~\bibnamefont {Zhong}},
  \bibinfo {author} {\bibfnamefont {J.}~\bibnamefont {Peng}}, \bibinfo {author}
  {\bibfnamefont {F.}~\bibnamefont {Li}}, \bibinfo {author} {\bibfnamefont
  {C.}~\bibnamefont {Nie}}, \bibinfo {author} {\bibfnamefont {L.}~\bibnamefont
  {Wang}}, \bibinfo {author} {\bibfnamefont {X.~J.}\ \bibnamefont {Zhou}},
  \bibinfo {author} {\bibfnamefont {X.}~\bibnamefont {Ma}}, \bibinfo {author}
  {\bibfnamefont {Q.~K.}\ \bibnamefont {Xue}}, \ and\ \bibinfo {author}
  {\bibfnamefont {F.}~\bibnamefont {Liu}},\ }\href
  {https://doi.org/10.1038/nmat4686} {\bibfield  {journal} {\bibinfo  {journal}
  {Nature Materials}\ }\textbf {\bibinfo {volume} {15}},\ \bibinfo {pages}
  {968} (\bibinfo {year} {2016})},\ \bibinfo {note} {article}\BibitemShut
  {NoStop}%
\bibitem [{\citenamefont {Jiang}\ \emph {et~al.}(2009)\citenamefont {Jiang},
  \citenamefont {Cheng}, \citenamefont {Sun},\ and\ \citenamefont
  {Xie}}]{Jiang2009}%
  \BibitemOpen
  \bibfield  {author} {\bibinfo {author} {\bibfnamefont {H.}~\bibnamefont
  {Jiang}}, \bibinfo {author} {\bibfnamefont {S.}~\bibnamefont {Cheng}},
  \bibinfo {author} {\bibfnamefont {Q.-f.}\ \bibnamefont {Sun}}, \ and\
  \bibinfo {author} {\bibfnamefont {X.~C.}\ \bibnamefont {Xie}},\ }\href
  {\doibase 10.1103/PhysRevLett.103.036803} {\bibfield  {journal} {\bibinfo
  {journal} {Phys. Rev. Lett.}\ }\textbf {\bibinfo {volume} {103}},\ \bibinfo
  {pages} {036803} (\bibinfo {year} {2009})}\BibitemShut {NoStop}%
\bibitem [{\citenamefont {Knez}\ \emph {et~al.}(2014)\citenamefont {Knez},
  \citenamefont {Rettner}, \citenamefont {Yang}, \citenamefont {Parkin},
  \citenamefont {Du}, \citenamefont {Du},\ and\ \citenamefont
  {Sullivan}}]{Ivan2014}%
  \BibitemOpen
  \bibfield  {author} {\bibinfo {author} {\bibfnamefont {I.}~\bibnamefont
  {Knez}}, \bibinfo {author} {\bibfnamefont {C.~T.}\ \bibnamefont {Rettner}},
  \bibinfo {author} {\bibfnamefont {S.-H.}\ \bibnamefont {Yang}}, \bibinfo
  {author} {\bibfnamefont {S.~S.~P.}\ \bibnamefont {Parkin}}, \bibinfo {author}
  {\bibfnamefont {L.}~\bibnamefont {Du}}, \bibinfo {author} {\bibfnamefont
  {R.-R.}\ \bibnamefont {Du}}, \ and\ \bibinfo {author} {\bibfnamefont
  {G.}~\bibnamefont {Sullivan}},\ }\href {\doibase
  10.1103/PhysRevLett.112.026602} {\bibfield  {journal} {\bibinfo  {journal}
  {Phys. Rev. Lett.}\ }\textbf {\bibinfo {volume} {112}},\ \bibinfo {pages}
  {026602} (\bibinfo {year} {2014})}\BibitemShut {NoStop}%
\bibitem [{\citenamefont {Haldane}(1988)}]{Haldane1988}%
  \BibitemOpen
  \bibfield  {author} {\bibinfo {author} {\bibfnamefont {F.~D.~M.}\
  \bibnamefont {Haldane}},\ }\href {\doibase 10.1103/PhysRevLett.61.2015}
  {\bibfield  {journal} {\bibinfo  {journal} {Phys. Rev. Lett.}\ }\textbf
  {\bibinfo {volume} {61}},\ \bibinfo {pages} {2015} (\bibinfo {year}
  {1988})}\BibitemShut {NoStop}%
\bibitem [{\citenamefont {Yu}\ \emph {et~al.}(2010)\citenamefont {Yu},
  \citenamefont {Zhang}, \citenamefont {Zhang}, \citenamefont {Zhang},
  \citenamefont {Dai},\ and\ \citenamefont {Fang}}]{Yurui2010}%
  \BibitemOpen
  \bibfield  {author} {\bibinfo {author} {\bibfnamefont {R.}~\bibnamefont
  {Yu}}, \bibinfo {author} {\bibfnamefont {W.}~\bibnamefont {Zhang}}, \bibinfo
  {author} {\bibfnamefont {H.-J.}\ \bibnamefont {Zhang}}, \bibinfo {author}
  {\bibfnamefont {S.-C.}\ \bibnamefont {Zhang}}, \bibinfo {author}
  {\bibfnamefont {X.}~\bibnamefont {Dai}}, \ and\ \bibinfo {author}
  {\bibfnamefont {Z.}~\bibnamefont {Fang}},\ }\href {\doibase
  10.1126/science.1187485} {\bibfield  {journal} {\bibinfo  {journal}
  {Science}\ }\textbf {\bibinfo {volume} {329}},\ \bibinfo {pages} {61}
  (\bibinfo {year} {2010})}\BibitemShut {NoStop}%
\bibitem [{\citenamefont {Chang}\ \emph {et~al.}(2013)\citenamefont {Chang},
  \citenamefont {Zhang}, \citenamefont {Feng}, \citenamefont {Shen},
  \citenamefont {Zhang}, \citenamefont {Guo}, \citenamefont {Li}, \citenamefont
  {Ou}, \citenamefont {Wei}, \citenamefont {Wang}, \citenamefont {Ji},
  \citenamefont {Feng}, \citenamefont {Ji}, \citenamefont {Chen}, \citenamefont
  {Jia}, \citenamefont {Dai}, \citenamefont {Fang}, \citenamefont {Zhang},
  \citenamefont {He}, \citenamefont {Wang}, \citenamefont {Lu}, \citenamefont
  {Ma},\ and\ \citenamefont {Xue}}]{Chang2013}%
  \BibitemOpen
  \bibfield  {author} {\bibinfo {author} {\bibfnamefont {C.-Z.}\ \bibnamefont
  {Chang}}, \bibinfo {author} {\bibfnamefont {J.}~\bibnamefont {Zhang}},
  \bibinfo {author} {\bibfnamefont {X.}~\bibnamefont {Feng}}, \bibinfo {author}
  {\bibfnamefont {J.}~\bibnamefont {Shen}}, \bibinfo {author} {\bibfnamefont
  {Z.}~\bibnamefont {Zhang}}, \bibinfo {author} {\bibfnamefont
  {M.}~\bibnamefont {Guo}}, \bibinfo {author} {\bibfnamefont {K.}~\bibnamefont
  {Li}}, \bibinfo {author} {\bibfnamefont {Y.}~\bibnamefont {Ou}}, \bibinfo
  {author} {\bibfnamefont {P.}~\bibnamefont {Wei}}, \bibinfo {author}
  {\bibfnamefont {L.-L.}\ \bibnamefont {Wang}}, \bibinfo {author}
  {\bibfnamefont {Z.-Q.}\ \bibnamefont {Ji}}, \bibinfo {author} {\bibfnamefont
  {Y.}~\bibnamefont {Feng}}, \bibinfo {author} {\bibfnamefont {S.}~\bibnamefont
  {Ji}}, \bibinfo {author} {\bibfnamefont {X.}~\bibnamefont {Chen}}, \bibinfo
  {author} {\bibfnamefont {J.}~\bibnamefont {Jia}}, \bibinfo {author}
  {\bibfnamefont {X.}~\bibnamefont {Dai}}, \bibinfo {author} {\bibfnamefont
  {Z.}~\bibnamefont {Fang}}, \bibinfo {author} {\bibfnamefont {S.-C.}\
  \bibnamefont {Zhang}}, \bibinfo {author} {\bibfnamefont {K.}~\bibnamefont
  {He}}, \bibinfo {author} {\bibfnamefont {Y.}~\bibnamefont {Wang}}, \bibinfo
  {author} {\bibfnamefont {L.}~\bibnamefont {Lu}}, \bibinfo {author}
  {\bibfnamefont {X.-C.}\ \bibnamefont {Ma}}, \ and\ \bibinfo {author}
  {\bibfnamefont {Q.-K.}\ \bibnamefont {Xue}},\ }\href {\doibase
  10.1126/science.1234414} {\bibfield  {journal} {\bibinfo  {journal}
  {Science}\ }\textbf {\bibinfo {volume} {340}},\ \bibinfo {pages} {167}
  (\bibinfo {year} {2013})}\BibitemShut {NoStop}%
\bibitem [{\citenamefont {Liu}\ \emph {et~al.}(2016)\citenamefont {Liu},
  \citenamefont {Zhang},\ and\ \citenamefont {Qi}}]{Liu2016}%
  \BibitemOpen
  \bibfield  {author} {\bibinfo {author} {\bibfnamefont {C.-X.}\ \bibnamefont
  {Liu}}, \bibinfo {author} {\bibfnamefont {S.-C.}\ \bibnamefont {Zhang}}, \
  and\ \bibinfo {author} {\bibfnamefont {X.-L.}\ \bibnamefont {Qi}},\ }\href
  {\doibase 10.1146/annurev-conmatphys-031115-011417} {\bibfield  {journal}
  {\bibinfo  {journal} {Annual Review of Condensed Matter Physics}\ }\textbf
  {\bibinfo {volume} {7}},\ \bibinfo {pages} {301} (\bibinfo {year}
  {2016})}\BibitemShut {NoStop}%
\bibitem [{\citenamefont {He}\ \emph {et~al.}(2018)\citenamefont {He},
  \citenamefont {Wang},\ and\ \citenamefont {Xue}}]{Heke2018}%
  \BibitemOpen
  \bibfield  {author} {\bibinfo {author} {\bibfnamefont {K.}~\bibnamefont
  {He}}, \bibinfo {author} {\bibfnamefont {Y.}~\bibnamefont {Wang}}, \ and\
  \bibinfo {author} {\bibfnamefont {Q.-K.}\ \bibnamefont {Xue}},\ }\href
  {\doibase 10.1146/annurev-conmatphys-033117-054144} {\bibfield  {journal}
  {\bibinfo  {journal} {Annual Review of Condensed Matter Physics}\ }\textbf
  {\bibinfo {volume} {9}},\ \bibinfo {pages} {329} (\bibinfo {year}
  {2018})}\BibitemShut {NoStop}%
\bibitem [{\citenamefont {Mogi}\ \emph
  {et~al.}(2017{\natexlab{a}})\citenamefont {Mogi}, \citenamefont {Kawamura},
  \citenamefont {Yoshimi}, \citenamefont {Tsukazaki}, \citenamefont {Kozuka},
  \citenamefont {Shirakawa}, \citenamefont {Takahashi}, \citenamefont
  {Kawasaki},\ and\ \citenamefont {Tokura}}]{Mogi2017}%
  \BibitemOpen
  \bibfield  {author} {\bibinfo {author} {\bibfnamefont {M.}~\bibnamefont
  {Mogi}}, \bibinfo {author} {\bibfnamefont {M.}~\bibnamefont {Kawamura}},
  \bibinfo {author} {\bibfnamefont {R.}~\bibnamefont {Yoshimi}}, \bibinfo
  {author} {\bibfnamefont {A.}~\bibnamefont {Tsukazaki}}, \bibinfo {author}
  {\bibfnamefont {Y.}~\bibnamefont {Kozuka}}, \bibinfo {author} {\bibfnamefont
  {N.}~\bibnamefont {Shirakawa}}, \bibinfo {author} {\bibfnamefont {K.~S.}\
  \bibnamefont {Takahashi}}, \bibinfo {author} {\bibfnamefont {M.}~\bibnamefont
  {Kawasaki}}, \ and\ \bibinfo {author} {\bibfnamefont {Y.}~\bibnamefont
  {Tokura}},\ }\href {https://doi.org/10.1038/nmat4855} {\bibfield  {journal}
  {\bibinfo  {journal} {Nature Materials}\ }\textbf {\bibinfo {volume} {16}},\
  \bibinfo {pages} {516 EP } (\bibinfo {year}
  {2017}{\natexlab{a}})}\BibitemShut {NoStop}%
\bibitem [{\citenamefont {Mogi}\ \emph
  {et~al.}(2017{\natexlab{b}})\citenamefont {Mogi}, \citenamefont {Kawamura},
  \citenamefont {Tsukazaki}, \citenamefont {Yoshimi}, \citenamefont
  {Takahashi}, \citenamefont {Kawasaki},\ and\ \citenamefont
  {Tokura}}]{Mogi2017b}%
  \BibitemOpen
  \bibfield  {author} {\bibinfo {author} {\bibfnamefont {M.}~\bibnamefont
  {Mogi}}, \bibinfo {author} {\bibfnamefont {M.}~\bibnamefont {Kawamura}},
  \bibinfo {author} {\bibfnamefont {A.}~\bibnamefont {Tsukazaki}}, \bibinfo
  {author} {\bibfnamefont {R.}~\bibnamefont {Yoshimi}}, \bibinfo {author}
  {\bibfnamefont {K.~S.}\ \bibnamefont {Takahashi}}, \bibinfo {author}
  {\bibfnamefont {M.}~\bibnamefont {Kawasaki}}, \ and\ \bibinfo {author}
  {\bibfnamefont {Y.}~\bibnamefont {Tokura}},\ }\href {\doibase
  10.1126/sciadv.aao1669} {\bibfield  {journal} {\bibinfo  {journal} {Science
  Advances}\ }\textbf {\bibinfo {volume} {3}} (\bibinfo {year}
  {2017}{\natexlab{b}}),\ 10.1126/sciadv.aao1669}\BibitemShut {NoStop}%
\bibitem [{\citenamefont {Xiao}\ \emph {et~al.}(2018)\citenamefont {Xiao},
  \citenamefont {Jiang}, \citenamefont {Shin}, \citenamefont {Wang},
  \citenamefont {Wang}, \citenamefont {Zhao}, \citenamefont {Liu},
  \citenamefont {Wu}, \citenamefont {Chan}, \citenamefont {Samarth},\ and\
  \citenamefont {Chang}}]{Chang2018}%
  \BibitemOpen
  \bibfield  {author} {\bibinfo {author} {\bibfnamefont {D.}~\bibnamefont
  {Xiao}}, \bibinfo {author} {\bibfnamefont {J.}~\bibnamefont {Jiang}},
  \bibinfo {author} {\bibfnamefont {J.-H.}\ \bibnamefont {Shin}}, \bibinfo
  {author} {\bibfnamefont {W.}~\bibnamefont {Wang}}, \bibinfo {author}
  {\bibfnamefont {F.}~\bibnamefont {Wang}}, \bibinfo {author} {\bibfnamefont
  {Y.-F.}\ \bibnamefont {Zhao}}, \bibinfo {author} {\bibfnamefont
  {C.}~\bibnamefont {Liu}}, \bibinfo {author} {\bibfnamefont {W.}~\bibnamefont
  {Wu}}, \bibinfo {author} {\bibfnamefont {M.~H.~W.}\ \bibnamefont {Chan}},
  \bibinfo {author} {\bibfnamefont {N.}~\bibnamefont {Samarth}}, \ and\
  \bibinfo {author} {\bibfnamefont {C.-Z.}\ \bibnamefont {Chang}},\ }\href
  {\doibase 10.1103/PhysRevLett.120.056801} {\bibfield  {journal} {\bibinfo
  {journal} {Phys. Rev. Lett.}\ }\textbf {\bibinfo {volume} {120}},\ \bibinfo
  {pages} {056801} (\bibinfo {year} {2018})}\BibitemShut {NoStop}%
\bibitem [{\citenamefont {Benalcazar}\ \emph
  {et~al.}(2017{\natexlab{a}})\citenamefont {Benalcazar}, \citenamefont
  {Bernevig},\ and\ \citenamefont {Hughes}}]{Benalcazar2017}%
  \BibitemOpen
  \bibfield  {author} {\bibinfo {author} {\bibfnamefont {W.~A.}\ \bibnamefont
  {Benalcazar}}, \bibinfo {author} {\bibfnamefont {B.~A.}\ \bibnamefont
  {Bernevig}}, \ and\ \bibinfo {author} {\bibfnamefont {T.~L.}\ \bibnamefont
  {Hughes}},\ }\href {\doibase 10.1126/science.aah6442} {\ \textbf {\bibinfo
  {volume} {357}},\ \bibinfo {pages} {61} (\bibinfo {year}
  {2017}{\natexlab{a}})}\BibitemShut {NoStop}%
\bibitem [{\citenamefont {Langbehn}\ \emph {et~al.}(2017)\citenamefont
  {Langbehn}, \citenamefont {Peng}, \citenamefont {Trifunovic}, \citenamefont
  {von Oppen},\ and\ \citenamefont {Brouwer}}]{Langbehn2017}%
  \BibitemOpen
  \bibfield  {author} {\bibinfo {author} {\bibfnamefont {J.}~\bibnamefont
  {Langbehn}}, \bibinfo {author} {\bibfnamefont {Y.}~\bibnamefont {Peng}},
  \bibinfo {author} {\bibfnamefont {L.}~\bibnamefont {Trifunovic}}, \bibinfo
  {author} {\bibfnamefont {F.}~\bibnamefont {von Oppen}}, \ and\ \bibinfo
  {author} {\bibfnamefont {P.~W.}\ \bibnamefont {Brouwer}},\ }\href {\doibase
  10.1103/PhysRevLett.119.246401} {\bibfield  {journal} {\bibinfo  {journal}
  {Phys. Rev. Lett.}\ }\textbf {\bibinfo {volume} {119}},\ \bibinfo {pages}
  {246401} (\bibinfo {year} {2017})}\BibitemShut {NoStop}%
\bibitem [{\citenamefont {Benalcazar}\ \emph
  {et~al.}(2017{\natexlab{b}})\citenamefont {Benalcazar}, \citenamefont
  {Bernevig},\ and\ \citenamefont {Hughes}}]{Benalcazar2017B}%
  \BibitemOpen
  \bibfield  {author} {\bibinfo {author} {\bibfnamefont {W.~A.}\ \bibnamefont
  {Benalcazar}}, \bibinfo {author} {\bibfnamefont {B.~A.}\ \bibnamefont
  {Bernevig}}, \ and\ \bibinfo {author} {\bibfnamefont {T.~L.}\ \bibnamefont
  {Hughes}},\ }\href {\doibase 10.1103/PhysRevB.96.245115} {\bibfield
  {journal} {\bibinfo  {journal} {Phys. Rev. B}\ }\textbf {\bibinfo {volume}
  {96}},\ \bibinfo {pages} {245115} (\bibinfo {year}
  {2017}{\natexlab{b}})}\BibitemShut {NoStop}%
\bibitem [{\citenamefont {Song}\ \emph {et~al.}(2017)\citenamefont {Song},
  \citenamefont {Fang},\ and\ \citenamefont {Fang}}]{Song2017}%
  \BibitemOpen
  \bibfield  {author} {\bibinfo {author} {\bibfnamefont {Z.}~\bibnamefont
  {Song}}, \bibinfo {author} {\bibfnamefont {Z.}~\bibnamefont {Fang}}, \ and\
  \bibinfo {author} {\bibfnamefont {C.}~\bibnamefont {Fang}},\ }\href {\doibase
  10.1103/PhysRevLett.119.246402} {\bibfield  {journal} {\bibinfo  {journal}
  {Phys. Rev. Lett.}\ }\textbf {\bibinfo {volume} {119}},\ \bibinfo {pages}
  {246402} (\bibinfo {year} {2017})}\BibitemShut {NoStop}%
\bibitem [{\citenamefont {Ezawa}(2018{\natexlab{a}})}]{Ezawa2018}%
  \BibitemOpen
  \bibfield  {author} {\bibinfo {author} {\bibfnamefont {M.}~\bibnamefont
  {Ezawa}},\ }\href {\doibase 10.1103/PhysRevLett.120.026801} {\bibfield
  {journal} {\bibinfo  {journal} {Phys. Rev. Lett.}\ }\textbf {\bibinfo
  {volume} {120}},\ \bibinfo {pages} {026801} (\bibinfo {year}
  {2018}{\natexlab{a}})}\BibitemShut {NoStop}%
\bibitem [{\citenamefont {Ezawa}(2018{\natexlab{b}})}]{Ezawa2018B}%
  \BibitemOpen
  \bibfield  {author} {\bibinfo {author} {\bibfnamefont {M.}~\bibnamefont
  {Ezawa}},\ }\href {\doibase 10.1103/PhysRevLett.121.116801} {\bibfield
  {journal} {\bibinfo  {journal} {Phys. Rev. Lett.}\ }\textbf {\bibinfo
  {volume} {121}},\ \bibinfo {pages} {116801} (\bibinfo {year}
  {2018}{\natexlab{b}})}\BibitemShut {NoStop}%
\bibitem [{\citenamefont {Yue}\ \emph {et~al.}(2019)\citenamefont {Yue},
  \citenamefont {Xu}, \citenamefont {Song}, \citenamefont {Weng}, \citenamefont
  {Lu}, \citenamefont {Fang},\ and\ \citenamefont {Dai}}]{Yue2019}%
  \BibitemOpen
  \bibfield  {author} {\bibinfo {author} {\bibfnamefont {C.}~\bibnamefont
  {Yue}}, \bibinfo {author} {\bibfnamefont {Y.}~\bibnamefont {Xu}}, \bibinfo
  {author} {\bibfnamefont {Z.}~\bibnamefont {Song}}, \bibinfo {author}
  {\bibfnamefont {H.}~\bibnamefont {Weng}}, \bibinfo {author} {\bibfnamefont
  {Y.-M.}\ \bibnamefont {Lu}}, \bibinfo {author} {\bibfnamefont
  {C.}~\bibnamefont {Fang}}, \ and\ \bibinfo {author} {\bibfnamefont
  {X.}~\bibnamefont {Dai}},\ }\href {\doibase 10.1038/s41567-019-0457-0}
  {\bibfield  {journal} {\bibinfo  {journal} {Nature Physics}\ }\textbf
  {\bibinfo {volume} {15}},\ \bibinfo {pages} {577¨C581} (\bibinfo {year}
  {2019})}\BibitemShut {NoStop}%
\bibitem [{\citenamefont {Baireuther}\ \emph {et~al.}(2014)\citenamefont
  {Baireuther}, \citenamefont {Edge}, \citenamefont {Fulga}, \citenamefont
  {Beenakker},\ and\ \citenamefont {Tworzyd\l{}o}}]{Baireuther2014}%
  \BibitemOpen
  \bibfield  {author} {\bibinfo {author} {\bibfnamefont {P.}~\bibnamefont
  {Baireuther}}, \bibinfo {author} {\bibfnamefont {J.~M.}\ \bibnamefont
  {Edge}}, \bibinfo {author} {\bibfnamefont {I.~C.}\ \bibnamefont {Fulga}},
  \bibinfo {author} {\bibfnamefont {C.~W.~J.}\ \bibnamefont {Beenakker}}, \
  and\ \bibinfo {author} {\bibfnamefont {J.}~\bibnamefont {Tworzyd\l{}o}},\
  }\href {\doibase 10.1103/PhysRevB.89.035410} {\bibfield  {journal} {\bibinfo
  {journal} {Phys. Rev. B}\ }\textbf {\bibinfo {volume} {89}},\ \bibinfo
  {pages} {035410} (\bibinfo {year} {2014})}\BibitemShut {NoStop}%
\bibitem [{\citenamefont {Schindler}\ \emph
  {et~al.}(2018{\natexlab{a}})\citenamefont {Schindler}, \citenamefont {Cook},
  \citenamefont {Vergniory}, \citenamefont {Wang}, \citenamefont {Parkin},
  \citenamefont {Bernevig},\ and\ \citenamefont {Neupert}}]{Schindler2018}%
  \BibitemOpen
  \bibfield  {author} {\bibinfo {author} {\bibfnamefont {F.}~\bibnamefont
  {Schindler}}, \bibinfo {author} {\bibfnamefont {A.~M.}\ \bibnamefont {Cook}},
  \bibinfo {author} {\bibfnamefont {M.~G.}\ \bibnamefont {Vergniory}}, \bibinfo
  {author} {\bibfnamefont {Z.}~\bibnamefont {Wang}}, \bibinfo {author}
  {\bibfnamefont {S.~S.~P.}\ \bibnamefont {Parkin}}, \bibinfo {author}
  {\bibfnamefont {B.~A.}\ \bibnamefont {Bernevig}}, \ and\ \bibinfo {author}
  {\bibfnamefont {T.}~\bibnamefont {Neupert}},\ }\href {\doibase
  10.1126/sciadv.aat0346} {\ \textbf {\bibinfo {volume} {4}} (\bibinfo {year}
  {2018}{\natexlab{a}}),\ 10.1126/sciadv.aat0346}\BibitemShut {NoStop}%
\bibitem [{\citenamefont {Schindler}\ \emph
  {et~al.}(2018{\natexlab{b}})\citenamefont {Schindler}, \citenamefont {Wang},
  \citenamefont {Vergniory}, \citenamefont {Cook}, \citenamefont {Murani},
  \citenamefont {Sengupta}, \citenamefont {Kasumov}, \citenamefont {Deblock},
  \citenamefont {Jeon}, \citenamefont {Drozdov}, \citenamefont {Bouchiat},
  \citenamefont {Gu{\'e}ron}, \citenamefont {Yazdani}, \citenamefont
  {Bernevig},\ and\ \citenamefont {Neupert}}]{Schindler2018B}%
  \BibitemOpen
  \bibfield  {author} {\bibinfo {author} {\bibfnamefont {F.}~\bibnamefont
  {Schindler}}, \bibinfo {author} {\bibfnamefont {Z.}~\bibnamefont {Wang}},
  \bibinfo {author} {\bibfnamefont {M.~G.}\ \bibnamefont {Vergniory}}, \bibinfo
  {author} {\bibfnamefont {A.~M.}\ \bibnamefont {Cook}}, \bibinfo {author}
  {\bibfnamefont {A.}~\bibnamefont {Murani}}, \bibinfo {author} {\bibfnamefont
  {S.}~\bibnamefont {Sengupta}}, \bibinfo {author} {\bibfnamefont {A.~Y.}\
  \bibnamefont {Kasumov}}, \bibinfo {author} {\bibfnamefont {R.}~\bibnamefont
  {Deblock}}, \bibinfo {author} {\bibfnamefont {S.}~\bibnamefont {Jeon}},
  \bibinfo {author} {\bibfnamefont {I.}~\bibnamefont {Drozdov}}, \bibinfo
  {author} {\bibfnamefont {H.}~\bibnamefont {Bouchiat}}, \bibinfo {author}
  {\bibfnamefont {S.}~\bibnamefont {Gu{\'e}ron}}, \bibinfo {author}
  {\bibfnamefont {A.}~\bibnamefont {Yazdani}}, \bibinfo {author} {\bibfnamefont
  {B.~A.}\ \bibnamefont {Bernevig}}, \ and\ \bibinfo {author} {\bibfnamefont
  {T.}~\bibnamefont {Neupert}},\ }\href {\doibase 10.1038/s41567-018-0224-7}
  {\bibfield  {journal} {\bibinfo  {journal} {Nature Physics}\ }\textbf
  {\bibinfo {volume} {14}},\ \bibinfo {pages} {918} (\bibinfo {year}
  {2018}{\natexlab{b}})}\BibitemShut {NoStop}%
\bibitem [{\citenamefont {Liu}\ \emph {et~al.}(2019{\natexlab{a}})\citenamefont
  {Liu}, \citenamefont {Deng},\ and\ \citenamefont {Wakabayashi}}]{Liu2019}%
  \BibitemOpen
  \bibfield  {author} {\bibinfo {author} {\bibfnamefont {F.}~\bibnamefont
  {Liu}}, \bibinfo {author} {\bibfnamefont {H.-Y.}\ \bibnamefont {Deng}}, \
  and\ \bibinfo {author} {\bibfnamefont {K.}~\bibnamefont {Wakabayashi}},\
  }\href {\doibase 10.1103/PhysRevLett.122.086804} {\bibfield  {journal}
  {\bibinfo  {journal} {Phys. Rev. Lett.}\ }\textbf {\bibinfo {volume} {122}},\
  \bibinfo {pages} {086804} (\bibinfo {year} {2019}{\natexlab{a}})}\BibitemShut
  {NoStop}%
\bibitem [{\citenamefont {Jiang}\ \emph {et~al.}(2018)\citenamefont {Jiang},
  \citenamefont {Feng}, \citenamefont {Wu}, \citenamefont {Li}, \citenamefont
  {Bai}, \citenamefont {Li}, \citenamefont {Zhang}, \citenamefont {Gu},
  \citenamefont {Feng}, \citenamefont {Zhang},\ and\ \citenamefont
  {et~al.}}]{MBE2018}%
  \BibitemOpen
  \bibfield  {author} {\bibinfo {author} {\bibfnamefont {G.}~\bibnamefont
  {Jiang}}, \bibinfo {author} {\bibfnamefont {Y.}~\bibnamefont {Feng}},
  \bibinfo {author} {\bibfnamefont {W.}~\bibnamefont {Wu}}, \bibinfo {author}
  {\bibfnamefont {S.}~\bibnamefont {Li}}, \bibinfo {author} {\bibfnamefont
  {Y.}~\bibnamefont {Bai}}, \bibinfo {author} {\bibfnamefont {Y.}~\bibnamefont
  {Li}}, \bibinfo {author} {\bibfnamefont {Q.}~\bibnamefont {Zhang}}, \bibinfo
  {author} {\bibfnamefont {L.}~\bibnamefont {Gu}}, \bibinfo {author}
  {\bibfnamefont {X.}~\bibnamefont {Feng}}, \bibinfo {author} {\bibfnamefont
  {D.}~\bibnamefont {Zhang}}, \ and\ \bibinfo {author} {\bibnamefont
  {et~al.}},\ }\href {\doibase 10.1088/0256-307x/35/7/076802} {\bibfield
  {journal} {\bibinfo  {journal} {Chinese Physics Letters}\ }\textbf {\bibinfo
  {volume} {35}},\ \bibinfo {pages} {076802} (\bibinfo {year}
  {2018})}\BibitemShut {NoStop}%
\bibitem [{\citenamefont {Otrokov}\ \emph {et~al.}(2018)\citenamefont
  {Otrokov}, \citenamefont {Klimovskikh}, \citenamefont {Bentmann},
  \citenamefont {Zeugner}, \citenamefont {Aliev}, \citenamefont {Gass},
  \citenamefont {Wolter}, \citenamefont {Koroleva}, \citenamefont {Estyunin},
  \citenamefont {Shikin}, \citenamefont {Blanco-Rey}, \citenamefont {Hoffmann},
  \citenamefont {Vyazovskaya}, \citenamefont {Eremeev}, \citenamefont
  {Koroteev}, \citenamefont {Amiraslanov}, \citenamefont {Babanly},
  \citenamefont {Mamedov}, \citenamefont {Abdullayev}, \citenamefont {Zverev},
  \citenamefont {B¨¹chner}, \citenamefont {Schwier}, \citenamefont {Kumar},
  \citenamefont {Kimura}, \citenamefont {Petaccia}, \citenamefont {Santo},
  \citenamefont {Vidal}, \citenamefont {Schatz}, \citenamefont {Ki{\ss}ner},
  \citenamefont {Min}, \citenamefont {Moser}, \citenamefont {Peixoto},
  \citenamefont {Reinert}, \citenamefont {Ernst}, \citenamefont {Echenique},
  \citenamefont {Isaeva},\ and\ \citenamefont {Chulkov}}]{otrokov2018}%
  \BibitemOpen
  \bibfield  {author} {\bibinfo {author} {\bibfnamefont {M.~M.}\ \bibnamefont
  {Otrokov}}, \bibinfo {author} {\bibfnamefont {I.~I.}\ \bibnamefont
  {Klimovskikh}}, \bibinfo {author} {\bibfnamefont {H.}~\bibnamefont
  {Bentmann}}, \bibinfo {author} {\bibfnamefont {A.}~\bibnamefont {Zeugner}},
  \bibinfo {author} {\bibfnamefont {Z.~S.}\ \bibnamefont {Aliev}}, \bibinfo
  {author} {\bibfnamefont {S.}~\bibnamefont {Gass}}, \bibinfo {author}
  {\bibfnamefont {A.~U.~B.}\ \bibnamefont {Wolter}}, \bibinfo {author}
  {\bibfnamefont {A.~V.}\ \bibnamefont {Koroleva}}, \bibinfo {author}
  {\bibfnamefont {D.}~\bibnamefont {Estyunin}}, \bibinfo {author}
  {\bibfnamefont {A.~M.}\ \bibnamefont {Shikin}}, \bibinfo {author}
  {\bibfnamefont {M.}~\bibnamefont {Blanco-Rey}}, \bibinfo {author}
  {\bibfnamefont {M.}~\bibnamefont {Hoffmann}}, \bibinfo {author}
  {\bibfnamefont {A.~Y.}\ \bibnamefont {Vyazovskaya}}, \bibinfo {author}
  {\bibfnamefont {S.~V.}\ \bibnamefont {Eremeev}}, \bibinfo {author}
  {\bibfnamefont {Y.~M.}\ \bibnamefont {Koroteev}}, \bibinfo {author}
  {\bibfnamefont {I.~R.}\ \bibnamefont {Amiraslanov}}, \bibinfo {author}
  {\bibfnamefont {M.~B.}\ \bibnamefont {Babanly}}, \bibinfo {author}
  {\bibfnamefont {N.~T.}\ \bibnamefont {Mamedov}}, \bibinfo {author}
  {\bibfnamefont {N.~A.}\ \bibnamefont {Abdullayev}}, \bibinfo {author}
  {\bibfnamefont {V.~N.}\ \bibnamefont {Zverev}}, \bibinfo {author}
  {\bibfnamefont {B.}~\bibnamefont {B¨¹chner}}, \bibinfo {author}
  {\bibfnamefont {E.~F.}\ \bibnamefont {Schwier}}, \bibinfo {author}
  {\bibfnamefont {S.}~\bibnamefont {Kumar}}, \bibinfo {author} {\bibfnamefont
  {A.}~\bibnamefont {Kimura}}, \bibinfo {author} {\bibfnamefont
  {L.}~\bibnamefont {Petaccia}}, \bibinfo {author} {\bibfnamefont {G.~D.}\
  \bibnamefont {Santo}}, \bibinfo {author} {\bibfnamefont {R.~C.}\ \bibnamefont
  {Vidal}}, \bibinfo {author} {\bibfnamefont {S.}~\bibnamefont {Schatz}},
  \bibinfo {author} {\bibfnamefont {K.}~\bibnamefont {Ki{\ss}ner}}, \bibinfo
  {author} {\bibfnamefont {C.-H.}\ \bibnamefont {Min}}, \bibinfo {author}
  {\bibfnamefont {S.~K.}\ \bibnamefont {Moser}}, \bibinfo {author}
  {\bibfnamefont {T.~R.~F.}\ \bibnamefont {Peixoto}}, \bibinfo {author}
  {\bibfnamefont {F.}~\bibnamefont {Reinert}}, \bibinfo {author} {\bibfnamefont
  {A.}~\bibnamefont {Ernst}}, \bibinfo {author} {\bibfnamefont {P.~M.}\
  \bibnamefont {Echenique}}, \bibinfo {author} {\bibfnamefont {A.}~\bibnamefont
  {Isaeva}}, \ and\ \bibinfo {author} {\bibfnamefont {E.~V.}\ \bibnamefont
  {Chulkov}},\ }\href@noop {} {\  (\bibinfo {year} {2018})},\ \Eprint
  {http://arxiv.org/abs/1809.07389} {arXiv:1809.07389 [cond-mat.mtrl-sci]}
  \BibitemShut {NoStop}%
\bibitem [{\citenamefont {Gong}\ \emph {et~al.}(2019)\citenamefont {Gong},
  \citenamefont {Guo}, \citenamefont {Li}, \citenamefont {Zhu}, \citenamefont
  {Liao}, \citenamefont {Liu}, \citenamefont {Zhang}, \citenamefont {Gu},
  \citenamefont {Tang}, \citenamefont {Feng}, \citenamefont {Zhang},
  \citenamefont {Li}, \citenamefont {Song}, \citenamefont {Wang}, \citenamefont
  {Yu}, \citenamefont {Chen}, \citenamefont {Wang}, \citenamefont {Yao},
  \citenamefont {Duan}, \citenamefont {Xu}, \citenamefont {Zhang},
  \citenamefont {Ma}, \citenamefont {Xue},\ and\ \citenamefont
  {He}}]{Heke2019}%
  \BibitemOpen
  \bibfield  {author} {\bibinfo {author} {\bibfnamefont {Y.}~\bibnamefont
  {Gong}}, \bibinfo {author} {\bibfnamefont {J.}~\bibnamefont {Guo}}, \bibinfo
  {author} {\bibfnamefont {J.}~\bibnamefont {Li}}, \bibinfo {author}
  {\bibfnamefont {K.}~\bibnamefont {Zhu}}, \bibinfo {author} {\bibfnamefont
  {M.}~\bibnamefont {Liao}}, \bibinfo {author} {\bibfnamefont {X.}~\bibnamefont
  {Liu}}, \bibinfo {author} {\bibfnamefont {Q.}~\bibnamefont {Zhang}}, \bibinfo
  {author} {\bibfnamefont {L.}~\bibnamefont {Gu}}, \bibinfo {author}
  {\bibfnamefont {L.}~\bibnamefont {Tang}}, \bibinfo {author} {\bibfnamefont
  {X.}~\bibnamefont {Feng}}, \bibinfo {author} {\bibfnamefont {D.}~\bibnamefont
  {Zhang}}, \bibinfo {author} {\bibfnamefont {W.}~\bibnamefont {Li}}, \bibinfo
  {author} {\bibfnamefont {C.}~\bibnamefont {Song}}, \bibinfo {author}
  {\bibfnamefont {L.}~\bibnamefont {Wang}}, \bibinfo {author} {\bibfnamefont
  {P.}~\bibnamefont {Yu}}, \bibinfo {author} {\bibfnamefont {X.}~\bibnamefont
  {Chen}}, \bibinfo {author} {\bibfnamefont {Y.}~\bibnamefont {Wang}}, \bibinfo
  {author} {\bibfnamefont {H.}~\bibnamefont {Yao}}, \bibinfo {author}
  {\bibfnamefont {W.}~\bibnamefont {Duan}}, \bibinfo {author} {\bibfnamefont
  {Y.}~\bibnamefont {Xu}}, \bibinfo {author} {\bibfnamefont {S.-C.}\
  \bibnamefont {Zhang}}, \bibinfo {author} {\bibfnamefont {X.}~\bibnamefont
  {Ma}}, \bibinfo {author} {\bibfnamefont {Q.-K.}\ \bibnamefont {Xue}}, \ and\
  \bibinfo {author} {\bibfnamefont {K.}~\bibnamefont {He}},\ }\href {\doibase
  10.1088/0256-307x/36/7/076801} {\bibfield  {journal} {\bibinfo  {journal}
  {Chinese Physics Letters}\ }\textbf {\bibinfo {volume} {36}},\ \bibinfo
  {pages} {076801} (\bibinfo {year} {2019})}\BibitemShut {NoStop}%
\bibitem [{\citenamefont {Li}\ \emph {et~al.}(2019{\natexlab{a}})\citenamefont
  {Li}, \citenamefont {Li}, \citenamefont {Du}, \citenamefont {Wang},
  \citenamefont {Gu}, \citenamefont {Zhang}, \citenamefont {He}, \citenamefont
  {Duan},\ and\ \citenamefont {Xu}}]{Lieaaw5685}%
  \BibitemOpen
  \bibfield  {author} {\bibinfo {author} {\bibfnamefont {J.}~\bibnamefont
  {Li}}, \bibinfo {author} {\bibfnamefont {Y.}~\bibnamefont {Li}}, \bibinfo
  {author} {\bibfnamefont {S.}~\bibnamefont {Du}}, \bibinfo {author}
  {\bibfnamefont {Z.}~\bibnamefont {Wang}}, \bibinfo {author} {\bibfnamefont
  {B.-L.}\ \bibnamefont {Gu}}, \bibinfo {author} {\bibfnamefont {S.-C.}\
  \bibnamefont {Zhang}}, \bibinfo {author} {\bibfnamefont {K.}~\bibnamefont
  {He}}, \bibinfo {author} {\bibfnamefont {W.}~\bibnamefont {Duan}}, \ and\
  \bibinfo {author} {\bibfnamefont {Y.}~\bibnamefont {Xu}},\ }\href {\doibase
  10.1126/sciadv.aaw5685} {\ \textbf {\bibinfo {volume} {5}} (\bibinfo {year}
  {2019}{\natexlab{a}}),\ 10.1126/sciadv.aaw5685}\BibitemShut {NoStop}%
\bibitem [{\citenamefont {Vidal}\ \emph
  {et~al.}(2019{\natexlab{a}})\citenamefont {Vidal}, \citenamefont {Bentmann},
  \citenamefont {Peixoto}, \citenamefont {Zeugner}, \citenamefont {Moser},
  \citenamefont {Min}, \citenamefont {Schatz}, \citenamefont {Ki\ss{}ner},
  \citenamefont {\"Unzelmann}, \citenamefont {Fornari}, \citenamefont {Vasili},
  \citenamefont {Valvidares}, \citenamefont {Sakamoto}, \citenamefont {Mondal},
  \citenamefont {Fujii}, \citenamefont {Vobornik}, \citenamefont {Jung},
  \citenamefont {Cacho}, \citenamefont {Kim}, \citenamefont {Koch},
  \citenamefont {Jozwiak}, \citenamefont {Bostwick}, \citenamefont {Denlinger},
  \citenamefont {Rotenberg}, \citenamefont {Buck}, \citenamefont {Hoesch},
  \citenamefont {Diekmann}, \citenamefont {Rohlf}, \citenamefont {Kall\"ane},
  \citenamefont {Rossnagel}, \citenamefont {Otrokov}, \citenamefont {Chulkov},
  \citenamefont {Ruck}, \citenamefont {Isaeva},\ and\ \citenamefont
  {Reinert}}]{Vidal2018}%
  \BibitemOpen
  \bibfield  {author} {\bibinfo {author} {\bibfnamefont {R.~C.}\ \bibnamefont
  {Vidal}}, \bibinfo {author} {\bibfnamefont {H.}~\bibnamefont {Bentmann}},
  \bibinfo {author} {\bibfnamefont {T.~R.~F.}\ \bibnamefont {Peixoto}},
  \bibinfo {author} {\bibfnamefont {A.}~\bibnamefont {Zeugner}}, \bibinfo
  {author} {\bibfnamefont {S.}~\bibnamefont {Moser}}, \bibinfo {author}
  {\bibfnamefont {C.-H.}\ \bibnamefont {Min}}, \bibinfo {author} {\bibfnamefont
  {S.}~\bibnamefont {Schatz}}, \bibinfo {author} {\bibfnamefont
  {K.}~\bibnamefont {Ki\ss{}ner}}, \bibinfo {author} {\bibfnamefont
  {M.}~\bibnamefont {\"Unzelmann}}, \bibinfo {author} {\bibfnamefont {C.~I.}\
  \bibnamefont {Fornari}}, \bibinfo {author} {\bibfnamefont {H.~B.}\
  \bibnamefont {Vasili}}, \bibinfo {author} {\bibfnamefont {M.}~\bibnamefont
  {Valvidares}}, \bibinfo {author} {\bibfnamefont {K.}~\bibnamefont
  {Sakamoto}}, \bibinfo {author} {\bibfnamefont {D.}~\bibnamefont {Mondal}},
  \bibinfo {author} {\bibfnamefont {J.}~\bibnamefont {Fujii}}, \bibinfo
  {author} {\bibfnamefont {I.}~\bibnamefont {Vobornik}}, \bibinfo {author}
  {\bibfnamefont {S.}~\bibnamefont {Jung}}, \bibinfo {author} {\bibfnamefont
  {C.}~\bibnamefont {Cacho}}, \bibinfo {author} {\bibfnamefont {T.~K.}\
  \bibnamefont {Kim}}, \bibinfo {author} {\bibfnamefont {R.~J.}\ \bibnamefont
  {Koch}}, \bibinfo {author} {\bibfnamefont {C.}~\bibnamefont {Jozwiak}},
  \bibinfo {author} {\bibfnamefont {A.}~\bibnamefont {Bostwick}}, \bibinfo
  {author} {\bibfnamefont {J.~D.}\ \bibnamefont {Denlinger}}, \bibinfo {author}
  {\bibfnamefont {E.}~\bibnamefont {Rotenberg}}, \bibinfo {author}
  {\bibfnamefont {J.}~\bibnamefont {Buck}}, \bibinfo {author} {\bibfnamefont
  {M.}~\bibnamefont {Hoesch}}, \bibinfo {author} {\bibfnamefont
  {F.}~\bibnamefont {Diekmann}}, \bibinfo {author} {\bibfnamefont
  {S.}~\bibnamefont {Rohlf}}, \bibinfo {author} {\bibfnamefont
  {M.}~\bibnamefont {Kall\"ane}}, \bibinfo {author} {\bibfnamefont
  {K.}~\bibnamefont {Rossnagel}}, \bibinfo {author} {\bibfnamefont {M.~M.}\
  \bibnamefont {Otrokov}}, \bibinfo {author} {\bibfnamefont {E.~V.}\
  \bibnamefont {Chulkov}}, \bibinfo {author} {\bibfnamefont {M.}~\bibnamefont
  {Ruck}}, \bibinfo {author} {\bibfnamefont {A.}~\bibnamefont {Isaeva}}, \ and\
  \bibinfo {author} {\bibfnamefont {F.}~\bibnamefont {Reinert}},\ }\href
  {\doibase 10.1103/PhysRevB.100.121104} {\bibfield  {journal} {\bibinfo
  {journal} {Phys. Rev. B}\ }\textbf {\bibinfo {volume} {100}},\ \bibinfo
  {pages} {121104} (\bibinfo {year} {2019}{\natexlab{a}})}\BibitemShut
  {NoStop}%
\bibitem [{\citenamefont {Hu}\ \emph {et~al.}(2019{\natexlab{a}})\citenamefont
  {Hu}, \citenamefont {Zhou}, \citenamefont {Liu}, \citenamefont {Liu},
  \citenamefont {Hao}, \citenamefont {Emmanouilidou}, \citenamefont {Sun},
  \citenamefont {Liu}, \citenamefont {Brawer}, \citenamefont {Ramirez},
  \citenamefont {Cao}, \citenamefont {Liu}, \citenamefont {Dessau},\ and\
  \citenamefont {Ni}}]{hu2019van}%
  \BibitemOpen
  \bibfield  {author} {\bibinfo {author} {\bibfnamefont {C.}~\bibnamefont
  {Hu}}, \bibinfo {author} {\bibfnamefont {X.}~\bibnamefont {Zhou}}, \bibinfo
  {author} {\bibfnamefont {P.}~\bibnamefont {Liu}}, \bibinfo {author}
  {\bibfnamefont {J.}~\bibnamefont {Liu}}, \bibinfo {author} {\bibfnamefont
  {P.}~\bibnamefont {Hao}}, \bibinfo {author} {\bibfnamefont {E.}~\bibnamefont
  {Emmanouilidou}}, \bibinfo {author} {\bibfnamefont {H.}~\bibnamefont {Sun}},
  \bibinfo {author} {\bibfnamefont {Y.}~\bibnamefont {Liu}}, \bibinfo {author}
  {\bibfnamefont {H.}~\bibnamefont {Brawer}}, \bibinfo {author} {\bibfnamefont
  {A.~P.}\ \bibnamefont {Ramirez}}, \bibinfo {author} {\bibfnamefont
  {H.}~\bibnamefont {Cao}}, \bibinfo {author} {\bibfnamefont {Q.}~\bibnamefont
  {Liu}}, \bibinfo {author} {\bibfnamefont {D.}~\bibnamefont {Dessau}}, \ and\
  \bibinfo {author} {\bibfnamefont {N.}~\bibnamefont {Ni}},\ }\href@noop {} {\
  (\bibinfo {year} {2019}{\natexlab{a}})},\ \Eprint
  {http://arxiv.org/abs/1905.02154} {arXiv:1905.02154 [cond-mat.mtrl-sci]}
  \BibitemShut {NoStop}%
\bibitem [{\citenamefont {Wu}\ \emph {et~al.}(2019)\citenamefont {Wu},
  \citenamefont {Liu}, \citenamefont {Sasase}, \citenamefont {Ienaga},
  \citenamefont {Obata}, \citenamefont {Yukawa}, \citenamefont {Horiba},
  \citenamefont {Kumigashira}, \citenamefont {Okuma}, \citenamefont
  {Inoshita},\ and\ \citenamefont {Hosono}}]{wu2019}%
  \BibitemOpen
  \bibfield  {author} {\bibinfo {author} {\bibfnamefont {J.}~\bibnamefont
  {Wu}}, \bibinfo {author} {\bibfnamefont {F.}~\bibnamefont {Liu}}, \bibinfo
  {author} {\bibfnamefont {M.}~\bibnamefont {Sasase}}, \bibinfo {author}
  {\bibfnamefont {K.}~\bibnamefont {Ienaga}}, \bibinfo {author} {\bibfnamefont
  {Y.}~\bibnamefont {Obata}}, \bibinfo {author} {\bibfnamefont
  {R.}~\bibnamefont {Yukawa}}, \bibinfo {author} {\bibfnamefont
  {K.}~\bibnamefont {Horiba}}, \bibinfo {author} {\bibfnamefont
  {H.}~\bibnamefont {Kumigashira}}, \bibinfo {author} {\bibfnamefont
  {S.}~\bibnamefont {Okuma}}, \bibinfo {author} {\bibfnamefont
  {T.}~\bibnamefont {Inoshita}}, \ and\ \bibinfo {author} {\bibfnamefont
  {H.}~\bibnamefont {Hosono}},\ }\href@noop {} {\  (\bibinfo {year} {2019})},\
  \Eprint {http://arxiv.org/abs/1905.02385} {arXiv:1905.02385
  [cond-mat.mtrl-sci]} \BibitemShut {NoStop}%
\bibitem [{\citenamefont {Li}\ \emph {et~al.}(2019{\natexlab{b}})\citenamefont
  {Li}, \citenamefont {Gao}, \citenamefont {Duan}, \citenamefont {Xu},
  \citenamefont {Zhu}, \citenamefont {Tian}, \citenamefont {Fan}, \citenamefont
  {Rao}, \citenamefont {Huang}, \citenamefont {Li}, \citenamefont {Liu},
  \citenamefont {Liu}, \citenamefont {Huang}, \citenamefont {Li}, \citenamefont
  {Liu}, \citenamefont {Zhang}, \citenamefont {Lei}, \citenamefont {Shi},
  \citenamefont {Zhang}, \citenamefont {Weng}, \citenamefont {Qian},\ and\
  \citenamefont {Ding}}]{li2019dirac}%
  \BibitemOpen
  \bibfield  {author} {\bibinfo {author} {\bibfnamefont {H.}~\bibnamefont
  {Li}}, \bibinfo {author} {\bibfnamefont {S.-Y.}\ \bibnamefont {Gao}},
  \bibinfo {author} {\bibfnamefont {S.-F.}\ \bibnamefont {Duan}}, \bibinfo
  {author} {\bibfnamefont {Y.-F.}\ \bibnamefont {Xu}}, \bibinfo {author}
  {\bibfnamefont {K.-J.}\ \bibnamefont {Zhu}}, \bibinfo {author} {\bibfnamefont
  {S.-J.}\ \bibnamefont {Tian}}, \bibinfo {author} {\bibfnamefont {W.-H.}\
  \bibnamefont {Fan}}, \bibinfo {author} {\bibfnamefont {Z.-C.}\ \bibnamefont
  {Rao}}, \bibinfo {author} {\bibfnamefont {J.-R.}\ \bibnamefont {Huang}},
  \bibinfo {author} {\bibfnamefont {J.-J.}\ \bibnamefont {Li}}, \bibinfo
  {author} {\bibfnamefont {Z.-T.}\ \bibnamefont {Liu}}, \bibinfo {author}
  {\bibfnamefont {W.-L.}\ \bibnamefont {Liu}}, \bibinfo {author} {\bibfnamefont
  {Y.-B.}\ \bibnamefont {Huang}}, \bibinfo {author} {\bibfnamefont {Y.-L.}\
  \bibnamefont {Li}}, \bibinfo {author} {\bibfnamefont {Y.}~\bibnamefont
  {Liu}}, \bibinfo {author} {\bibfnamefont {G.-B.}\ \bibnamefont {Zhang}},
  \bibinfo {author} {\bibfnamefont {H.-C.}\ \bibnamefont {Lei}}, \bibinfo
  {author} {\bibfnamefont {Y.-G.}\ \bibnamefont {Shi}}, \bibinfo {author}
  {\bibfnamefont {W.-T.}\ \bibnamefont {Zhang}}, \bibinfo {author}
  {\bibfnamefont {H.-M.}\ \bibnamefont {Weng}}, \bibinfo {author}
  {\bibfnamefont {T.}~\bibnamefont {Qian}}, \ and\ \bibinfo {author}
  {\bibfnamefont {H.}~\bibnamefont {Ding}},\ }\href@noop {} {\  (\bibinfo
  {year} {2019}{\natexlab{b}})},\ \Eprint {http://arxiv.org/abs/1907.06491}
  {arXiv:1907.06491 [cond-mat.mtrl-sci]} \BibitemShut {NoStop}%
\bibitem [{\citenamefont {Hao}\ \emph {et~al.}(2019)\citenamefont {Hao},
  \citenamefont {Liu}, \citenamefont {Feng}, \citenamefont {Ma}, \citenamefont
  {Schwier}, \citenamefont {Arita}, \citenamefont {Kumar}, \citenamefont {Hu},
  \citenamefont {Lu}, \citenamefont {Zeng}, \citenamefont {Wang}, \citenamefont
  {Hao}, \citenamefont {Sun}, \citenamefont {Zhang}, \citenamefont {Mei},
  \citenamefont {Ni}, \citenamefont {Wu}, \citenamefont {Shimada},
  \citenamefont {Chen}, \citenamefont {Liu},\ and\ \citenamefont
  {Liu}}]{hao2019gapless}%
  \BibitemOpen
  \bibfield  {author} {\bibinfo {author} {\bibfnamefont {Y.-J.}\ \bibnamefont
  {Hao}}, \bibinfo {author} {\bibfnamefont {P.}~\bibnamefont {Liu}}, \bibinfo
  {author} {\bibfnamefont {Y.}~\bibnamefont {Feng}}, \bibinfo {author}
  {\bibfnamefont {X.-M.}\ \bibnamefont {Ma}}, \bibinfo {author} {\bibfnamefont
  {E.~F.}\ \bibnamefont {Schwier}}, \bibinfo {author} {\bibfnamefont
  {M.}~\bibnamefont {Arita}}, \bibinfo {author} {\bibfnamefont
  {S.}~\bibnamefont {Kumar}}, \bibinfo {author} {\bibfnamefont
  {C.}~\bibnamefont {Hu}}, \bibinfo {author} {\bibfnamefont {R.}~\bibnamefont
  {Lu}}, \bibinfo {author} {\bibfnamefont {M.}~\bibnamefont {Zeng}}, \bibinfo
  {author} {\bibfnamefont {Y.}~\bibnamefont {Wang}}, \bibinfo {author}
  {\bibfnamefont {Z.}~\bibnamefont {Hao}}, \bibinfo {author} {\bibfnamefont
  {H.}~\bibnamefont {Sun}}, \bibinfo {author} {\bibfnamefont {K.}~\bibnamefont
  {Zhang}}, \bibinfo {author} {\bibfnamefont {J.}~\bibnamefont {Mei}}, \bibinfo
  {author} {\bibfnamefont {N.}~\bibnamefont {Ni}}, \bibinfo {author}
  {\bibfnamefont {L.}~\bibnamefont {Wu}}, \bibinfo {author} {\bibfnamefont
  {K.}~\bibnamefont {Shimada}}, \bibinfo {author} {\bibfnamefont
  {C.}~\bibnamefont {Chen}}, \bibinfo {author} {\bibfnamefont {Q.}~\bibnamefont
  {Liu}}, \ and\ \bibinfo {author} {\bibfnamefont {C.}~\bibnamefont {Liu}},\
  }\href@noop {} {\  (\bibinfo {year} {2019})},\ \Eprint
  {http://arxiv.org/abs/1907.03722} {arXiv:1907.03722 [cond-mat.mtrl-sci]}
  \BibitemShut {NoStop}%
\bibitem [{\citenamefont {Chen}\ \emph {et~al.}(2019)\citenamefont {Chen},
  \citenamefont {Xu}, \citenamefont {Li}, \citenamefont {Li}, \citenamefont
  {Zhang}, \citenamefont {Li}, \citenamefont {Wu}, \citenamefont {Liang},
  \citenamefont {Chen}, \citenamefont {Jung}, \citenamefont {Cacho},
  \citenamefont {Wang}, \citenamefont {Mao}, \citenamefont {Liu}, \citenamefont
  {Wang}, \citenamefont {Guo}, \citenamefont {Xu}, \citenamefont {Liu},
  \citenamefont {Yang},\ and\ \citenamefont {Chen}}]{chen2019topological}%
  \BibitemOpen
  \bibfield  {author} {\bibinfo {author} {\bibfnamefont {Y.~J.}\ \bibnamefont
  {Chen}}, \bibinfo {author} {\bibfnamefont {L.~X.}\ \bibnamefont {Xu}},
  \bibinfo {author} {\bibfnamefont {J.~H.}\ \bibnamefont {Li}}, \bibinfo
  {author} {\bibfnamefont {Y.~W.}\ \bibnamefont {Li}}, \bibinfo {author}
  {\bibfnamefont {C.~F.}\ \bibnamefont {Zhang}}, \bibinfo {author}
  {\bibfnamefont {H.}~\bibnamefont {Li}}, \bibinfo {author} {\bibfnamefont
  {Y.}~\bibnamefont {Wu}}, \bibinfo {author} {\bibfnamefont {A.~J.}\
  \bibnamefont {Liang}}, \bibinfo {author} {\bibfnamefont {C.}~\bibnamefont
  {Chen}}, \bibinfo {author} {\bibfnamefont {S.~W.}\ \bibnamefont {Jung}},
  \bibinfo {author} {\bibfnamefont {C.}~\bibnamefont {Cacho}}, \bibinfo
  {author} {\bibfnamefont {H.~Y.}\ \bibnamefont {Wang}}, \bibinfo {author}
  {\bibfnamefont {Y.~H.}\ \bibnamefont {Mao}}, \bibinfo {author} {\bibfnamefont
  {S.}~\bibnamefont {Liu}}, \bibinfo {author} {\bibfnamefont {M.~X.}\
  \bibnamefont {Wang}}, \bibinfo {author} {\bibfnamefont {Y.~F.}\ \bibnamefont
  {Guo}}, \bibinfo {author} {\bibfnamefont {Y.}~\bibnamefont {Xu}}, \bibinfo
  {author} {\bibfnamefont {Z.~K.}\ \bibnamefont {Liu}}, \bibinfo {author}
  {\bibfnamefont {L.~X.}\ \bibnamefont {Yang}}, \ and\ \bibinfo {author}
  {\bibfnamefont {Y.~L.}\ \bibnamefont {Chen}},\ }\href@noop {} {\  (\bibinfo
  {year} {2019})},\ \Eprint {http://arxiv.org/abs/1907.05119} {arXiv:1907.05119
  [cond-mat.mtrl-sci]} \BibitemShut {NoStop}%
\bibitem [{\citenamefont {Deng}\ \emph {et~al.}(2019)\citenamefont {Deng},
  \citenamefont {Yu}, \citenamefont {Shi}, \citenamefont {Wang}, \citenamefont
  {Chen},\ and\ \citenamefont {Zhang}}]{yuanbo2019}%
  \BibitemOpen
  \bibfield  {author} {\bibinfo {author} {\bibfnamefont {Y.}~\bibnamefont
  {Deng}}, \bibinfo {author} {\bibfnamefont {Y.}~\bibnamefont {Yu}}, \bibinfo
  {author} {\bibfnamefont {M.~Z.}\ \bibnamefont {Shi}}, \bibinfo {author}
  {\bibfnamefont {J.}~\bibnamefont {Wang}}, \bibinfo {author} {\bibfnamefont
  {X.~H.}\ \bibnamefont {Chen}}, \ and\ \bibinfo {author} {\bibfnamefont
  {Y.}~\bibnamefont {Zhang}},\ }\href@noop {} {\  (\bibinfo {year} {2019})},\
  \Eprint {http://arxiv.org/abs/1904.11468} {arXiv:1904.11468
  [cond-mat.mtrl-sci]} \BibitemShut {NoStop}%
\bibitem [{\citenamefont {Ge}\ \emph {et~al.}(2019)\citenamefont {Ge},
  \citenamefont {Liu}, \citenamefont {Li}, \citenamefont {Li}, \citenamefont
  {Luo}, \citenamefont {Wu}, \citenamefont {Xu},\ and\ \citenamefont
  {Wang}}]{wangjian2019}%
  \BibitemOpen
  \bibfield  {author} {\bibinfo {author} {\bibfnamefont {J.}~\bibnamefont
  {Ge}}, \bibinfo {author} {\bibfnamefont {Y.}~\bibnamefont {Liu}}, \bibinfo
  {author} {\bibfnamefont {J.}~\bibnamefont {Li}}, \bibinfo {author}
  {\bibfnamefont {H.}~\bibnamefont {Li}}, \bibinfo {author} {\bibfnamefont
  {T.}~\bibnamefont {Luo}}, \bibinfo {author} {\bibfnamefont {Y.}~\bibnamefont
  {Wu}}, \bibinfo {author} {\bibfnamefont {Y.}~\bibnamefont {Xu}}, \ and\
  \bibinfo {author} {\bibfnamefont {J.}~\bibnamefont {Wang}},\ }\href@noop {}
  {\  (\bibinfo {year} {2019})},\ \Eprint {http://arxiv.org/abs/1907.09947}
  {arXiv:1907.09947 [cond-mat.mes-hall]} \BibitemShut {NoStop}%
\bibitem [{\citenamefont {Liu}\ \emph {et~al.}(2019{\natexlab{b}})\citenamefont
  {Liu}, \citenamefont {Wang}, \citenamefont {Li}, \citenamefont {Wu},
  \citenamefont {Li}, \citenamefont {Li}, \citenamefont {He}, \citenamefont
  {Xu}, \citenamefont {Zhang},\ and\ \citenamefont {Wang}}]{Yayu2019}%
  \BibitemOpen
  \bibfield  {author} {\bibinfo {author} {\bibfnamefont {C.}~\bibnamefont
  {Liu}}, \bibinfo {author} {\bibfnamefont {Y.}~\bibnamefont {Wang}}, \bibinfo
  {author} {\bibfnamefont {H.}~\bibnamefont {Li}}, \bibinfo {author}
  {\bibfnamefont {Y.}~\bibnamefont {Wu}}, \bibinfo {author} {\bibfnamefont
  {Y.}~\bibnamefont {Li}}, \bibinfo {author} {\bibfnamefont {J.}~\bibnamefont
  {Li}}, \bibinfo {author} {\bibfnamefont {K.}~\bibnamefont {He}}, \bibinfo
  {author} {\bibfnamefont {Y.}~\bibnamefont {Xu}}, \bibinfo {author}
  {\bibfnamefont {J.}~\bibnamefont {Zhang}}, \ and\ \bibinfo {author}
  {\bibfnamefont {Y.}~\bibnamefont {Wang}},\ }\href@noop {} {\  (\bibinfo
  {year} {2019}{\natexlab{b}})},\ \Eprint {http://arxiv.org/abs/1905.00715}
  {arXiv:1905.00715 [cond-mat.mes-hall]} \BibitemShut {NoStop}%
\bibitem [{\citenamefont {Zhang}\ \emph {et~al.}(2019)\citenamefont {Zhang},
  \citenamefont {Shi}, \citenamefont {Zhu}, \citenamefont {Xing}, \citenamefont
  {Zhang},\ and\ \citenamefont {Wang}}]{Wangjing}%
  \BibitemOpen
  \bibfield  {author} {\bibinfo {author} {\bibfnamefont {D.}~\bibnamefont
  {Zhang}}, \bibinfo {author} {\bibfnamefont {M.}~\bibnamefont {Shi}}, \bibinfo
  {author} {\bibfnamefont {T.}~\bibnamefont {Zhu}}, \bibinfo {author}
  {\bibfnamefont {D.}~\bibnamefont {Xing}}, \bibinfo {author} {\bibfnamefont
  {H.}~\bibnamefont {Zhang}}, \ and\ \bibinfo {author} {\bibfnamefont
  {J.}~\bibnamefont {Wang}},\ }\href {\doibase 10.1103/PhysRevLett.122.206401}
  {\bibfield  {journal} {\bibinfo  {journal} {Phys. Rev. Lett.}\ }\textbf
  {\bibinfo {volume} {122}},\ \bibinfo {pages} {206401} (\bibinfo {year}
  {2019})}\BibitemShut {NoStop}%
\bibitem [{\citenamefont {Vidal}\ \emph
  {et~al.}(2019{\natexlab{b}})\citenamefont {Vidal}, \citenamefont {Zeugner},
  \citenamefont {Facio}, \citenamefont {Ray}, \citenamefont {Haghighi},
  \citenamefont {Wolter}, \citenamefont {Bohorquez}, \citenamefont {Caglieris},
  \citenamefont {Moser}, \citenamefont {Figgemeier}, \citenamefont {Peixoto},
  \citenamefont {Vasili}, \citenamefont {Valvidares}, \citenamefont {Jung},
  \citenamefont {Cacho}, \citenamefont {Alfonsov}, \citenamefont {Mehlawat},
  \citenamefont {Kataev}, \citenamefont {Hess}, \citenamefont {Richter},
  \citenamefont {B¨¹chner}, \citenamefont {van~den Brink}, \citenamefont
  {Ruck}, \citenamefont {Reinert}, \citenamefont {Bentmann},\ and\
  \citenamefont {Isaeva}}]{vidal2019topological}%
  \BibitemOpen
  \bibfield  {author} {\bibinfo {author} {\bibfnamefont {R.~C.}\ \bibnamefont
  {Vidal}}, \bibinfo {author} {\bibfnamefont {A.}~\bibnamefont {Zeugner}},
  \bibinfo {author} {\bibfnamefont {J.~I.}\ \bibnamefont {Facio}}, \bibinfo
  {author} {\bibfnamefont {R.}~\bibnamefont {Ray}}, \bibinfo {author}
  {\bibfnamefont {M.~H.}\ \bibnamefont {Haghighi}}, \bibinfo {author}
  {\bibfnamefont {A.~U.~B.}\ \bibnamefont {Wolter}}, \bibinfo {author}
  {\bibfnamefont {L.~T.~C.}\ \bibnamefont {Bohorquez}}, \bibinfo {author}
  {\bibfnamefont {F.}~\bibnamefont {Caglieris}}, \bibinfo {author}
  {\bibfnamefont {S.}~\bibnamefont {Moser}}, \bibinfo {author} {\bibfnamefont
  {T.}~\bibnamefont {Figgemeier}}, \bibinfo {author} {\bibfnamefont {T.~R.~F.}\
  \bibnamefont {Peixoto}}, \bibinfo {author} {\bibfnamefont {H.~B.}\
  \bibnamefont {Vasili}}, \bibinfo {author} {\bibfnamefont {M.}~\bibnamefont
  {Valvidares}}, \bibinfo {author} {\bibfnamefont {S.}~\bibnamefont {Jung}},
  \bibinfo {author} {\bibfnamefont {C.}~\bibnamefont {Cacho}}, \bibinfo
  {author} {\bibfnamefont {A.}~\bibnamefont {Alfonsov}}, \bibinfo {author}
  {\bibfnamefont {K.}~\bibnamefont {Mehlawat}}, \bibinfo {author}
  {\bibfnamefont {V.}~\bibnamefont {Kataev}}, \bibinfo {author} {\bibfnamefont
  {C.}~\bibnamefont {Hess}}, \bibinfo {author} {\bibfnamefont {M.}~\bibnamefont
  {Richter}}, \bibinfo {author} {\bibfnamefont {B.}~\bibnamefont {B¨¹chner}},
  \bibinfo {author} {\bibfnamefont {J.}~\bibnamefont {van~den Brink}}, \bibinfo
  {author} {\bibfnamefont {M.}~\bibnamefont {Ruck}}, \bibinfo {author}
  {\bibfnamefont {F.}~\bibnamefont {Reinert}}, \bibinfo {author} {\bibfnamefont
  {H.}~\bibnamefont {Bentmann}}, \ and\ \bibinfo {author} {\bibfnamefont
  {A.}~\bibnamefont {Isaeva}},\ }\href@noop {} {\  (\bibinfo {year}
  {2019}{\natexlab{b}})},\ \Eprint {http://arxiv.org/abs/1906.08394}
  {arXiv:1906.08394 [cond-mat.mtrl-sci]} \BibitemShut {NoStop}%
\bibitem [{\citenamefont {Shi}\ \emph {et~al.}(2019)\citenamefont {Shi},
  \citenamefont {Lei}, \citenamefont {Zhu}, \citenamefont {Ma}, \citenamefont
  {Cui}, \citenamefont {Sun}, \citenamefont {Ying},\ and\ \citenamefont
  {Chen}}]{Shi_2019}%
  \BibitemOpen
  \bibfield  {author} {\bibinfo {author} {\bibfnamefont {M.~Z.}\ \bibnamefont
  {Shi}}, \bibinfo {author} {\bibfnamefont {B.}~\bibnamefont {Lei}}, \bibinfo
  {author} {\bibfnamefont {C.~S.}\ \bibnamefont {Zhu}}, \bibinfo {author}
  {\bibfnamefont {D.~H.}\ \bibnamefont {Ma}}, \bibinfo {author} {\bibfnamefont
  {J.~H.}\ \bibnamefont {Cui}}, \bibinfo {author} {\bibfnamefont {Z.~L.}\
  \bibnamefont {Sun}}, \bibinfo {author} {\bibfnamefont {J.~J.}\ \bibnamefont
  {Ying}}, \ and\ \bibinfo {author} {\bibfnamefont {X.~H.}\ \bibnamefont
  {Chen}},\ }\href {\doibase 10.1103/PhysRevB.100.155144} {\bibfield  {journal}
  {\bibinfo  {journal} {Phys. Rev. B}\ }\textbf {\bibinfo {volume} {100}},\
  \bibinfo {pages} {155144} (\bibinfo {year} {2019})}\BibitemShut {NoStop}%
\bibitem [{\citenamefont {Ding}\ \emph {et~al.}(2019)\citenamefont {Ding},
  \citenamefont {Hu}, \citenamefont {Ye}, \citenamefont {Feng}, \citenamefont
  {Ni},\ and\ \citenamefont {Cao}}]{ding2019crystal}%
  \BibitemOpen
  \bibfield  {author} {\bibinfo {author} {\bibfnamefont {L.}~\bibnamefont
  {Ding}}, \bibinfo {author} {\bibfnamefont {C.}~\bibnamefont {Hu}}, \bibinfo
  {author} {\bibfnamefont {F.}~\bibnamefont {Ye}}, \bibinfo {author}
  {\bibfnamefont {E.}~\bibnamefont {Feng}}, \bibinfo {author} {\bibfnamefont
  {N.}~\bibnamefont {Ni}}, \ and\ \bibinfo {author} {\bibfnamefont
  {H.}~\bibnamefont {Cao}},\ }\href@noop {} {\  (\bibinfo {year} {2019})},\
  \Eprint {http://arxiv.org/abs/1910.06248} {arXiv:1910.06248
  [cond-mat.str-el]} \BibitemShut {NoStop}%
\bibitem [{\citenamefont {Hu}\ \emph {et~al.}(2019{\natexlab{b}})\citenamefont
  {Hu}, \citenamefont {Xu}, \citenamefont {Shi}, \citenamefont {Luo},
  \citenamefont {Peng}, \citenamefont {Wang}, \citenamefont {Ying},
  \citenamefont {Wu}, \citenamefont {Liu}, \citenamefont {Zhang}, \citenamefont
  {Chen}, \citenamefont {Xu}, \citenamefont {Chen},\ and\ \citenamefont
  {He}}]{hu2019universal}%
  \BibitemOpen
  \bibfield  {author} {\bibinfo {author} {\bibfnamefont {Y.}~\bibnamefont
  {Hu}}, \bibinfo {author} {\bibfnamefont {L.}~\bibnamefont {Xu}}, \bibinfo
  {author} {\bibfnamefont {M.}~\bibnamefont {Shi}}, \bibinfo {author}
  {\bibfnamefont {A.}~\bibnamefont {Luo}}, \bibinfo {author} {\bibfnamefont
  {S.}~\bibnamefont {Peng}}, \bibinfo {author} {\bibfnamefont {Z.~Y.}\
  \bibnamefont {Wang}}, \bibinfo {author} {\bibfnamefont {J.~J.}\ \bibnamefont
  {Ying}}, \bibinfo {author} {\bibfnamefont {T.}~\bibnamefont {Wu}}, \bibinfo
  {author} {\bibfnamefont {Z.~K.}\ \bibnamefont {Liu}}, \bibinfo {author}
  {\bibfnamefont {C.~F.}\ \bibnamefont {Zhang}}, \bibinfo {author}
  {\bibfnamefont {Y.~L.}\ \bibnamefont {Chen}}, \bibinfo {author}
  {\bibfnamefont {G.}~\bibnamefont {Xu}}, \bibinfo {author} {\bibfnamefont
  {X.~H.}\ \bibnamefont {Chen}}, \ and\ \bibinfo {author} {\bibfnamefont
  {J.~F.}\ \bibnamefont {He}},\ }\href@noop {} {\  (\bibinfo {year}
  {2019}{\natexlab{b}})},\ \Eprint {http://arxiv.org/abs/1910.11323}
  {arXiv:1910.11323 [cond-mat.mtrl-sci]} \BibitemShut {NoStop}%
\bibitem [{\citenamefont {Xu}\ \emph {et~al.}(2019)\citenamefont {Xu},
  \citenamefont {Mao}, \citenamefont {Wang}, \citenamefont {Li}, \citenamefont
  {Chen}, \citenamefont {Xia}, \citenamefont {Li}, \citenamefont {Zhang},
  \citenamefont {Zheng}, \citenamefont {Huang}, \citenamefont {Zhang},
  \citenamefont {Cui}, \citenamefont {Liang}, \citenamefont {Xia},
  \citenamefont {Su}, \citenamefont {Jung}, \citenamefont {Cacho},
  \citenamefont {Wang}, \citenamefont {Li}, \citenamefont {Xu}, \citenamefont
  {Guo}, \citenamefont {Yang}, \citenamefont {Liu},\ and\ \citenamefont
  {Chen}}]{xu2019persistent}%
  \BibitemOpen
  \bibfield  {author} {\bibinfo {author} {\bibfnamefont {L.~X.}\ \bibnamefont
  {Xu}}, \bibinfo {author} {\bibfnamefont {Y.~H.}\ \bibnamefont {Mao}},
  \bibinfo {author} {\bibfnamefont {H.~Y.}\ \bibnamefont {Wang}}, \bibinfo
  {author} {\bibfnamefont {J.~H.}\ \bibnamefont {Li}}, \bibinfo {author}
  {\bibfnamefont {Y.~J.}\ \bibnamefont {Chen}}, \bibinfo {author}
  {\bibfnamefont {Y.~Y.~Y.}\ \bibnamefont {Xia}}, \bibinfo {author}
  {\bibfnamefont {Y.~W.}\ \bibnamefont {Li}}, \bibinfo {author} {\bibfnamefont
  {J.}~\bibnamefont {Zhang}}, \bibinfo {author} {\bibfnamefont {H.~J.}\
  \bibnamefont {Zheng}}, \bibinfo {author} {\bibfnamefont {K.}~\bibnamefont
  {Huang}}, \bibinfo {author} {\bibfnamefont {C.~F.}\ \bibnamefont {Zhang}},
  \bibinfo {author} {\bibfnamefont {S.~T.}\ \bibnamefont {Cui}}, \bibinfo
  {author} {\bibfnamefont {A.~J.}\ \bibnamefont {Liang}}, \bibinfo {author}
  {\bibfnamefont {W.}~\bibnamefont {Xia}}, \bibinfo {author} {\bibfnamefont
  {H.}~\bibnamefont {Su}}, \bibinfo {author} {\bibfnamefont {S.~W.}\
  \bibnamefont {Jung}}, \bibinfo {author} {\bibfnamefont {C.}~\bibnamefont
  {Cacho}}, \bibinfo {author} {\bibfnamefont {M.~X.}\ \bibnamefont {Wang}},
  \bibinfo {author} {\bibfnamefont {G.}~\bibnamefont {Li}}, \bibinfo {author}
  {\bibfnamefont {Y.}~\bibnamefont {Xu}}, \bibinfo {author} {\bibfnamefont
  {Y.~F.}\ \bibnamefont {Guo}}, \bibinfo {author} {\bibfnamefont {L.~X.}\
  \bibnamefont {Yang}}, \bibinfo {author} {\bibfnamefont {Z.~K.}\ \bibnamefont
  {Liu}}, \ and\ \bibinfo {author} {\bibfnamefont {Y.~L.}\ \bibnamefont
  {Chen}},\ }\href@noop {} {\  (\bibinfo {year} {2019})},\ \Eprint
  {http://arxiv.org/abs/1910.11014} {arXiv:1910.11014 [cond-mat.mtrl-sci]}
  \BibitemShut {NoStop}%
\bibitem [{\citenamefont {Tian}\ \emph {et~al.}(2019)\citenamefont {Tian},
  \citenamefont {Gao}, \citenamefont {Nie}, \citenamefont {Qian}, \citenamefont
  {Gong}, \citenamefont {Fu}, \citenamefont {Li}, \citenamefont {Fan},
  \citenamefont {Zhang}, \citenamefont {Kondo}, \citenamefont {Shin},
  \citenamefont {Adell}, \citenamefont {Fedderwitz}, \citenamefont {Ding},
  \citenamefont {Wang}, \citenamefont {Qian},\ and\ \citenamefont
  {Lei}}]{tian2019magnetic}%
  \BibitemOpen
  \bibfield  {author} {\bibinfo {author} {\bibfnamefont {S.}~\bibnamefont
  {Tian}}, \bibinfo {author} {\bibfnamefont {S.}~\bibnamefont {Gao}}, \bibinfo
  {author} {\bibfnamefont {S.}~\bibnamefont {Nie}}, \bibinfo {author}
  {\bibfnamefont {Y.}~\bibnamefont {Qian}}, \bibinfo {author} {\bibfnamefont
  {C.}~\bibnamefont {Gong}}, \bibinfo {author} {\bibfnamefont {Y.}~\bibnamefont
  {Fu}}, \bibinfo {author} {\bibfnamefont {H.}~\bibnamefont {Li}}, \bibinfo
  {author} {\bibfnamefont {W.}~\bibnamefont {Fan}}, \bibinfo {author}
  {\bibfnamefont {P.}~\bibnamefont {Zhang}}, \bibinfo {author} {\bibfnamefont
  {T.}~\bibnamefont {Kondo}}, \bibinfo {author} {\bibfnamefont
  {S.}~\bibnamefont {Shin}}, \bibinfo {author} {\bibfnamefont {J.}~\bibnamefont
  {Adell}}, \bibinfo {author} {\bibfnamefont {H.}~\bibnamefont {Fedderwitz}},
  \bibinfo {author} {\bibfnamefont {H.}~\bibnamefont {Ding}}, \bibinfo {author}
  {\bibfnamefont {Z.}~\bibnamefont {Wang}}, \bibinfo {author} {\bibfnamefont
  {T.}~\bibnamefont {Qian}}, \ and\ \bibinfo {author} {\bibfnamefont
  {H.}~\bibnamefont {Lei}},\ }\href@noop {} {\  (\bibinfo {year} {2019})},\
  \Eprint {http://arxiv.org/abs/1910.10101} {arXiv:1910.10101
  [cond-mat.mtrl-sci]} \BibitemShut {NoStop}%
\bibitem [{\citenamefont {Du}\ \emph {et~al.}(2015)\citenamefont {Du},
  \citenamefont {Knez}, \citenamefont {Sullivan},\ and\ \citenamefont
  {Du}}]{Du2015}%
  \BibitemOpen
  \bibfield  {author} {\bibinfo {author} {\bibfnamefont {L.}~\bibnamefont
  {Du}}, \bibinfo {author} {\bibfnamefont {I.}~\bibnamefont {Knez}}, \bibinfo
  {author} {\bibfnamefont {G.}~\bibnamefont {Sullivan}}, \ and\ \bibinfo
  {author} {\bibfnamefont {R.-R.}\ \bibnamefont {Du}},\ }\href {\doibase
  10.1103/PhysRevLett.114.096802} {\bibfield  {journal} {\bibinfo  {journal}
  {Phys. Rev. Lett.}\ }\textbf {\bibinfo {volume} {114}},\ \bibinfo {pages}
  {096802} (\bibinfo {year} {2015})}\BibitemShut {NoStop}%
\bibitem [{\citenamefont {Wu}\ \emph {et~al.}(2018)\citenamefont {Wu},
  \citenamefont {Fatemi}, \citenamefont {Gibson}, \citenamefont {Watanabe},
  \citenamefont {Taniguchi}, \citenamefont {Cava},\ and\ \citenamefont
  {Jarillo-Herrero}}]{Wu2018}%
  \BibitemOpen
  \bibfield  {author} {\bibinfo {author} {\bibfnamefont {S.}~\bibnamefont
  {Wu}}, \bibinfo {author} {\bibfnamefont {V.}~\bibnamefont {Fatemi}}, \bibinfo
  {author} {\bibfnamefont {Q.~D.}\ \bibnamefont {Gibson}}, \bibinfo {author}
  {\bibfnamefont {K.}~\bibnamefont {Watanabe}}, \bibinfo {author}
  {\bibfnamefont {T.}~\bibnamefont {Taniguchi}}, \bibinfo {author}
  {\bibfnamefont {R.~J.}\ \bibnamefont {Cava}}, \ and\ \bibinfo {author}
  {\bibfnamefont {P.}~\bibnamefont {Jarillo-Herrero}},\ }\href {\doibase
  10.1126/science.aan6003} {\bibfield  {journal} {\bibinfo  {journal}
  {Science}\ }\textbf {\bibinfo {volume} {359}},\ \bibinfo {pages} {76}
  (\bibinfo {year} {2018})}\BibitemShut {NoStop}%
\bibitem [{\citenamefont {Su}\ \emph {et~al.}(1979)\citenamefont {Su},
  \citenamefont {Schrieffer},\ and\ \citenamefont {Heeger}}]{SSH1979}%
  \BibitemOpen
  \bibfield  {author} {\bibinfo {author} {\bibfnamefont {W.~P.}\ \bibnamefont
  {Su}}, \bibinfo {author} {\bibfnamefont {J.~R.}\ \bibnamefont {Schrieffer}},
  \ and\ \bibinfo {author} {\bibfnamefont {A.~J.}\ \bibnamefont {Heeger}},\
  }\href {\doibase 10.1103/PhysRevLett.42.1698} {\bibfield  {journal} {\bibinfo
   {journal} {Phys. Rev. Lett.}\ }\textbf {\bibinfo {volume} {42}},\ \bibinfo
  {pages} {1698} (\bibinfo {year} {1979})}\BibitemShut {NoStop}%
\bibitem [{\citenamefont {Rice}\ and\ \citenamefont {Mele}(1982)}]{RM1982}%
  \BibitemOpen
  \bibfield  {author} {\bibinfo {author} {\bibfnamefont {M.~J.}\ \bibnamefont
  {Rice}}\ and\ \bibinfo {author} {\bibfnamefont {E.~J.}\ \bibnamefont
  {Mele}},\ }\href {\doibase 10.1103/PhysRevLett.49.1455} {\bibfield  {journal}
  {\bibinfo  {journal} {Phys. Rev. Lett.}\ }\textbf {\bibinfo {volume} {49}},\
  \bibinfo {pages} {1455} (\bibinfo {year} {1982})}\BibitemShut {NoStop}%
\bibitem [{\citenamefont {Prodan}(2009)}]{Prodan2009}%
  \BibitemOpen
  \bibfield  {author} {\bibinfo {author} {\bibfnamefont {E.}~\bibnamefont
  {Prodan}},\ }\href {\doibase 10.1103/PhysRevB.80.125327} {\bibfield
  {journal} {\bibinfo  {journal} {Phys. Rev. B}\ }\textbf {\bibinfo {volume}
  {80}},\ \bibinfo {pages} {125327} (\bibinfo {year} {2009})}\BibitemShut
  {NoStop}%
\bibitem [{\citenamefont {Prodan}(2011)}]{Prodan2011}%
  \BibitemOpen
  \bibfield  {author} {\bibinfo {author} {\bibfnamefont {E.}~\bibnamefont
  {Prodan}},\ }\href {http://stacks.iop.org/1751-8121/44/i=11/a=113001}
  {\bibfield  {journal} {\bibinfo  {journal} {Journal of Physics A:
  Mathematical and Theoretical}\ }\textbf {\bibinfo {volume} {44}},\ \bibinfo
  {pages} {113001} (\bibinfo {year} {2011})}\BibitemShut {NoStop}%
\bibitem [{\citenamefont {Sheng}\ \emph {et~al.}(2005)\citenamefont {Sheng},
  \citenamefont {Sheng}, \citenamefont {Ting},\ and\ \citenamefont
  {Haldane}}]{Sheng2005}%
  \BibitemOpen
  \bibfield  {author} {\bibinfo {author} {\bibfnamefont {L.}~\bibnamefont
  {Sheng}}, \bibinfo {author} {\bibfnamefont {D.~N.}\ \bibnamefont {Sheng}},
  \bibinfo {author} {\bibfnamefont {C.~S.}\ \bibnamefont {Ting}}, \ and\
  \bibinfo {author} {\bibfnamefont {F.~D.~M.}\ \bibnamefont {Haldane}},\ }\href
  {\doibase 10.1103/PhysRevLett.95.136602} {\bibfield  {journal} {\bibinfo
  {journal} {Phys. Rev. Lett.}\ }\textbf {\bibinfo {volume} {95}},\ \bibinfo
  {pages} {136602} (\bibinfo {year} {2005})}\BibitemShut {NoStop}%
\bibitem [{\citenamefont {Sheng}\ \emph {et~al.}(2006)\citenamefont {Sheng},
  \citenamefont {Weng}, \citenamefont {Sheng},\ and\ \citenamefont
  {Haldane}}]{Sheng2006}%
  \BibitemOpen
  \bibfield  {author} {\bibinfo {author} {\bibfnamefont {D.~N.}\ \bibnamefont
  {Sheng}}, \bibinfo {author} {\bibfnamefont {Z.~Y.}\ \bibnamefont {Weng}},
  \bibinfo {author} {\bibfnamefont {L.}~\bibnamefont {Sheng}}, \ and\ \bibinfo
  {author} {\bibfnamefont {F.~D.~M.}\ \bibnamefont {Haldane}},\ }\href
  {\doibase 10.1103/PhysRevLett.97.036808} {\bibfield  {journal} {\bibinfo
  {journal} {Phys. Rev. Lett.}\ }\textbf {\bibinfo {volume} {97}},\ \bibinfo
  {pages} {036808} (\bibinfo {year} {2006})}\BibitemShut {NoStop}%
\bibitem [{\citenamefont {Datta}(1995)}]{datta1995}%
  \BibitemOpen
  \bibfield  {author} {\bibinfo {author} {\bibfnamefont {S.}~\bibnamefont
  {Datta}},\ }\href {\doibase 10.1017/CBO9780511805776} {\emph {\bibinfo
  {title} {Electronic Transport in Mesoscopic Systems}}},\ Cambridge Studies in
  Semiconductor Physics and Microelectronic Engineering\ (\bibinfo  {publisher}
  {Cambridge University Press},\ \bibinfo {year} {1995})\BibitemShut {NoStop}%
\bibitem [{\citenamefont {Jiang}\ \emph {et~al.}(2014)\citenamefont {Jiang},
  \citenamefont {Liu}, \citenamefont {Feng}, \citenamefont {Sun},\ and\
  \citenamefont {Xie}}]{Jiang2014}%
  \BibitemOpen
  \bibfield  {author} {\bibinfo {author} {\bibfnamefont {H.}~\bibnamefont
  {Jiang}}, \bibinfo {author} {\bibfnamefont {H.}~\bibnamefont {Liu}}, \bibinfo
  {author} {\bibfnamefont {J.}~\bibnamefont {Feng}}, \bibinfo {author}
  {\bibfnamefont {Q.}~\bibnamefont {Sun}}, \ and\ \bibinfo {author}
  {\bibfnamefont {X.~C.}\ \bibnamefont {Xie}},\ }\href {\doibase
  10.1103/PhysRevLett.112.176601} {\bibfield  {journal} {\bibinfo  {journal}
  {Phys. Rev. Lett.}\ }\textbf {\bibinfo {volume} {112}},\ \bibinfo {pages}
  {176601} (\bibinfo {year} {2014})}\BibitemShut {NoStop}%
\bibitem [{\citenamefont {B\"uttiker}(1986)}]{BM1986}%
  \BibitemOpen
  \bibfield  {author} {\bibinfo {author} {\bibfnamefont {M.}~\bibnamefont
  {B\"uttiker}},\ }\href {\doibase 10.1103/PhysRevB.33.3020} {\bibfield
  {journal} {\bibinfo  {journal} {Phys. Rev. B}\ }\textbf {\bibinfo {volume}
  {33}},\ \bibinfo {pages} {3020} (\bibinfo {year} {1986})}\BibitemShut
  {NoStop}%
\bibitem [{\citenamefont {Shi}\ and\ \citenamefont {Xie}(2001)}]{Shi2001}%
  \BibitemOpen
  \bibfield  {author} {\bibinfo {author} {\bibfnamefont {J.}~\bibnamefont
  {Shi}}\ and\ \bibinfo {author} {\bibfnamefont {X.~C.}\ \bibnamefont {Xie}},\
  }\href {\doibase 10.1103/PhysRevB.63.045123} {\bibfield  {journal} {\bibinfo
  {journal} {Phys. Rev. B}\ }\textbf {\bibinfo {volume} {63}},\ \bibinfo
  {pages} {045123} (\bibinfo {year} {2001})}\BibitemShut {NoStop}%
\bibitem [{\citenamefont {Xing}\ \emph {et~al.}(2008)\citenamefont {Xing},
  \citenamefont {Sun},\ and\ \citenamefont {Wang}}]{Xing2008}%
  \BibitemOpen
  \bibfield  {author} {\bibinfo {author} {\bibfnamefont {Y.}~\bibnamefont
  {Xing}}, \bibinfo {author} {\bibfnamefont {Q.-f.}\ \bibnamefont {Sun}}, \
  and\ \bibinfo {author} {\bibfnamefont {J.}~\bibnamefont {Wang}},\ }\href
  {\doibase 10.1103/PhysRevB.77.115346} {\bibfield  {journal} {\bibinfo
  {journal} {Phys. Rev. B}\ }\textbf {\bibinfo {volume} {77}},\ \bibinfo
  {pages} {115346} (\bibinfo {year} {2008})}\BibitemShut {NoStop}%
\bibitem [{\citenamefont {Ou}\ \emph {et~al.}(2018)\citenamefont {Ou},
  \citenamefont {Liu}, \citenamefont {Jiang}, \citenamefont {Feng},
  \citenamefont {Zhao}, \citenamefont {Wu}, \citenamefont {Wang}, \citenamefont
  {Li}, \citenamefont {Song}, \citenamefont {Wang}, \citenamefont {Wang},
  \citenamefont {Wu}, \citenamefont {Wang}, \citenamefont {He}, \citenamefont
  {Ma},\ and\ \citenamefont {Xue}}]{Yunbo2018}%
  \BibitemOpen
  \bibfield  {author} {\bibinfo {author} {\bibfnamefont {Y.}~\bibnamefont
  {Ou}}, \bibinfo {author} {\bibfnamefont {C.}~\bibnamefont {Liu}}, \bibinfo
  {author} {\bibfnamefont {G.}~\bibnamefont {Jiang}}, \bibinfo {author}
  {\bibfnamefont {Y.}~\bibnamefont {Feng}}, \bibinfo {author} {\bibfnamefont
  {D.}~\bibnamefont {Zhao}}, \bibinfo {author} {\bibfnamefont {W.}~\bibnamefont
  {Wu}}, \bibinfo {author} {\bibfnamefont {X.-X.}\ \bibnamefont {Wang}},
  \bibinfo {author} {\bibfnamefont {W.}~\bibnamefont {Li}}, \bibinfo {author}
  {\bibfnamefont {C.}~\bibnamefont {Song}}, \bibinfo {author} {\bibfnamefont
  {L.-L.}\ \bibnamefont {Wang}}, \bibinfo {author} {\bibfnamefont
  {W.}~\bibnamefont {Wang}}, \bibinfo {author} {\bibfnamefont {W.}~\bibnamefont
  {Wu}}, \bibinfo {author} {\bibfnamefont {Y.}~\bibnamefont {Wang}}, \bibinfo
  {author} {\bibfnamefont {K.}~\bibnamefont {He}}, \bibinfo {author}
  {\bibfnamefont {X.-C.}\ \bibnamefont {Ma}}, \ and\ \bibinfo {author}
  {\bibfnamefont {Q.-K.}\ \bibnamefont {Xue}},\ }\href {\doibase
  10.1002/adma.201703062} {\bibfield  {journal} {\bibinfo  {journal} {Advanced
  Materials}\ }\textbf {\bibinfo {volume} {30}},\ \bibinfo {pages} {1703062}
  (\bibinfo {year} {2018})}\BibitemShut {NoStop}%
\bibitem [{\citenamefont {Feng}\ \emph {et~al.}(2019)\citenamefont {Feng},
  \citenamefont {Jiang}, \citenamefont {Wu}, \citenamefont {Li}, \citenamefont
  {He}, \citenamefont {Ma}, \citenamefont {Xue},\ and\ \citenamefont
  {Wang}}]{Fengyang2019}%
  \BibitemOpen
  \bibfield  {author} {\bibinfo {author} {\bibfnamefont {Y.}~\bibnamefont
  {Feng}}, \bibinfo {author} {\bibfnamefont {G.}~\bibnamefont {Jiang}},
  \bibinfo {author} {\bibfnamefont {W.}~\bibnamefont {Wu}}, \bibinfo {author}
  {\bibfnamefont {S.}~\bibnamefont {Li}}, \bibinfo {author} {\bibfnamefont
  {K.}~\bibnamefont {He}}, \bibinfo {author} {\bibfnamefont {X.}~\bibnamefont
  {Ma}}, \bibinfo {author} {\bibfnamefont {Q.-K.}\ \bibnamefont {Xue}}, \ and\
  \bibinfo {author} {\bibfnamefont {Y.}~\bibnamefont {Wang}},\ }\href {\doibase
  10.1103/PhysRevB.100.165403} {\bibfield  {journal} {\bibinfo  {journal}
  {Phys. Rev. B}\ }\textbf {\bibinfo {volume} {100}},\ \bibinfo {pages}
  {165403} (\bibinfo {year} {2019})}\BibitemShut {NoStop}%
\bibitem [{\citenamefont {Khalaf}\ \emph {et~al.}(2019)\citenamefont {Khalaf},
  \citenamefont {Benalcazar}, \citenamefont {Hughes},\ and\ \citenamefont
  {Queiroz}}]{EK2019}%
  \BibitemOpen
  \bibfield  {author} {\bibinfo {author} {\bibfnamefont {E.}~\bibnamefont
  {Khalaf}}, \bibinfo {author} {\bibfnamefont {W.~A.}\ \bibnamefont
  {Benalcazar}}, \bibinfo {author} {\bibfnamefont {T.~L.}\ \bibnamefont
  {Hughes}}, \ and\ \bibinfo {author} {\bibfnamefont {R.}~\bibnamefont
  {Queiroz}},\ }\href@noop {} {\enquote {\bibinfo {title} {Boundary-obstructed
  topological phases},}\ } (\bibinfo {year} {2019}),\ \Eprint
  {http://arxiv.org/abs/1908.00011} {arXiv:1908.00011 [cond-mat.mes-hall]}
  \BibitemShut {NoStop}%
\end{thebibliography}
\end{document}